\documentclass[
 reprint,
 amsmath,amssymb,
 prd,
 superscriptaddress
]{revtex4-1}
\pdfoutput=1
\usepackage{graphicx}
\usepackage[utf8]{inputenc}
\usepackage{lineno}

\usepackage{color}
\usepackage{xcolor}

\begin{document}

\title{Darker and brighter branes, suppression and enhancement of photon production in a strongly coupled magnetized plasma}

\author{Daniel \'Avila}
 \email{davhdz06@ciencias.unam.mx}
\affiliation{Departamento de F\'isica, Facultad de Ciencias, Universidad Nacional Aut\'onoma de M\'exico, \\  A.P. 70-542, Ciudad de M\'exico 04510, M\'exico}
\affiliation{Departamento de Física de Altas Energias, Instituto de Ciencias Nucleares,
Universidad Nacional Autónoma de México Apartado Postal 70-543, CDMX 04510, México} 
\author{Francisco Nettel}
 \email{fnettel@ciencias.unam.mx}
 \affiliation{Departamento de F\'isica, Facultad de Ciencias, Universidad Nacional Aut\'onoma de M\'exico, \\  A.P. 70-542, Ciudad de M\'exico 04510, M\'exico}
\author{Leonardo Pati\~no}
 \email{leopj@ciencias.unam.mx}
\affiliation{Departamento de F\'isica, Facultad de Ciencias, Universidad Nacional Aut\'onoma de M\'exico, \\  A.P. 70-542, Ciudad de M\'exico 04510, M\'exico}


\begin{abstract}
We extend our holographic analysis of the emission of photons by a strongly coupled plasma subject to a very intense external magnetic field. We previously showed that in a particular model, any photon produced by the plasma had to be in its only polarization state parallel to the reaction plane. In this paper we consider a construction that relaxes a formerly imposed constraint, permitting the emission of photons with either out-plane or in-plane polarization. This constitutes a completion of our former study because the fully back-reacted equations decouple for these two polarization states in such a manner that those involving the in-plane are identical to the ones we explored previously. In view of the above, part of the details concerning the calculations and of the numerical results for the differential rate of emitted photons that we present here correspond exactly to those omitted in our preceding letter. One of our main results is that the production of photons is increased by the introduction of a non-vanishing magnetic field with an intensity up to a value $B_{\vartheta}$, above which the effect is reversed and said production becomes lower than the $B=0$ case. The characteristic intensity $B_{\vartheta}$ depends on the propagation direction and tends to zero as the photon momentum becomes aligned with the magnetic field. Additionally, we also show that the magnetic field has the effect of increasing the value of the elliptic flow, providing a possible explanation for the excess measured in collision experiments. The holographic model is constructed using an effective five-dimensional action that includes a scalar field in addition to the constant magnetic field.
\end{abstract}


\maketitle
\section{Introduction}
\label{Introduction}

One of the most important probes to obtain information about the quark-gluon plasma (QGP) are the thermal photons emitted by it. Due to the weakness of the electromagnetic coupling and the small extension in space of the QGP, the emitted photons exit the plasma practically unscattered, hence providing an excellent source of information about the emission point \cite{Arleo:2004gn,Wilde:2012wc,PHENIX:2011oxq,doi:10.1146/annurev.nucl.53.041002.110533,David:2019wpt}. It has been suggested that the QGP has global quark spin polarization in non-central heavy-ion collisions \cite{Liang:2004xn,Liang:2004ph}, and it was shown later that this in turn leads to the polarization of the emitted photons \cite{Ipp:2007ng}, either direct \cite{Baym:2014qfa} or virtual \cite{Baym:2017gzx,Baym:2017qxy,Speranza:2018osi}. 

Moreover, it has become increasingly accepted that an intense magnetic field pointing perpendicularly to the reaction plane is produced in high energy collisions \cite{Skokov:2009qp,Basar:2012bp,Andersen:2014xxa,Gursoy:2018yai,Tuchin:2013ie,Deng:2012pc} and that understanding its effects is relevant to properly analyze experimental observations \cite{Ayala:2018wux,Ayala:2020wzl,Ayala:2021lor,Ayala:2021nhx}. The maximum estimate for the intensity of said magnetic field was found in \cite{Deng:2012pc} to be in the range of $e B=5 m_{\pi}^{2}$ at $\sqrt{s}=200$ GeV in RHIC and $e B= 70 m_{\pi}^{2}$ at $\sqrt{s}=2.76$ TeV in LHC, with $m_{\pi}$ the mass of the neutral pion. These maxima are predicted to be respectively reached 1 fm/c and 0.4 fm/c after the corresponding collision, and decay approximately two orders of magnitude in about twice such times. It was suggested in \cite{Basar:2012bp} that the existence of strong (electro)magnetic fields in heavy ion collisions could account for the unexpected observations regarding the total emission of photons in general and the elliptic anisotropy of those with low transverse momentum $p_{\rm T}$, both underestimated by the methods of pQCD \cite{Wilde:2012wc,PHENIX:2011oxq}. Additionally, it was proposed in \cite{Yee:2013qma} that this magnetic field could lead to the quark spin polarization, and in turn induce a polarization on the emitted photons.

Given that the QGP produced at high energy p-p or heavy-ion collisions exists in a strongly coupled state \cite{Shuryak:2003xe,Shuryak:2004cy}, the gauge/gravity correspondence \cite{Maldacena:1997re} has been extensively used to explore some of its dynamical properties. It should be noted however, that the theory that can be studied with holographic methods is not properly quantum chromodynamics (QCD), as the exact gravity dual to this theory has not been found yet. What has been done instead is to consider theories similar to QCD, such as $\mathcal{N} = 4$ Super Yang-Mills (SYM) at finite temperature with gauge group $SU(N_{c})$, and either modify them to bring them as close to QCD as possible, or use them to compute quantities that are not sensitive to the details of the theory. An important example of the later is the famous shear viscosity to entropy ratio, which can be computed in the strongly coupled plasma of the SYM $\mathcal{N} = 4$ theory using holographic methods, and extrapolated to the QGP produced in high energy heavy-ion collisions \cite{Policastro:2001yc}. On the other hand, the modifications to SYM $\mathcal{N}=4$ have been extended to include spatial anisotropies \cite{Janik:2008tc,Mateos:2011ix,Mateos:2011tv} and, in particular, the presence of a very intense external magnetic field \cite{DHoker:2009mmn,Avila:2018hsi,Avila:2019pua,Avila:2020ved,Ammon:2020rvg}. 

The photon production of strongly coupled plasmas has been analyzed using many of the previously mentioned holographic models. While the first holographic study only considered SYM $\mathcal{N}=4$ at finite temperature \cite{CaronHuot:2006te}, many developments have been considered to improve the modelling of the experimental context. In \cite{Mateos:2007yp} $N_{f}$ flavor degrees of freedom were added to the theory in the probe limit, while the Veneziano limit was latter considered in \cite{Iatrakis:2016ugz}. Other modifications such as non-vanishing chemical potential \cite{Parnachev:2006ev,Jo:2010sg,Bu:2012zza} and spatial anisotropies \cite{Patino:2012py,Arefeva:2021jpa} have also been considered.

Of particular interest to this work is the inclusion of an external magnetic field. The photon production in this context was studied in \cite{Mamo:2013efa,Arciniega:2013dqa,Wu:2013qja} using the holographic model developed in \cite{DHoker:2009mmn}, finding that in general the magnetic field enhances it. The five-dimensional theory presented there, that considers the full backreaction of the magnetic field, is dual to the desired gauge theory because it is a solution to a consistent truncation of supergravity (SUGRA) IIB \cite{Cvetic:1999xp}. Given that the magnetic field is introduced by factorizing a $U(1)$ from the $SO(6)$ symmetry of the compact space and changing it to a gauge symmetry, from the gauge theory perspective the magnetic field in this case couples to the conserved current associated with a $U(1)$ subgroup of the $SU(4)$ $R-$symmetry. 

The effect of the magnetic field on the polarization of the emitted photon was studied in \cite{Yee:2013qma} by means of the Sakai-Sugimoto holographic model \cite{Sakai:2004cn}, where said magnetic field was introduced on the world volume of the $D8/\bar{D8}$-branes. However, the reported polarization was mild, given that the backreaction on the embedding of the branes resulting from the combined effect of the magnetic field and the electromagnetic perturbations was not taken into account.

Another important observable in the study of the emitted photons is the elliptic flow $v_{2}$, which characterizes the momentum anisotropy of the particles produced in heavy-ion collisions. Previous measurements of the elliptic flow at the RHIC and LHC collaborations revealed a surprisingly large value of $v_{2}$ for $p_{\rm T} < 4$ Gev/c \cite{PHENIX:2011oxq, Lohner:2012ct,David:2019wpt}, and the presence of an intense magnetic field presents itself as a possible explanation for such a large elliptic anisotropy at low $p_{\rm T}$ momenta. An holographic model for this was given in \cite{Yee:2013qma} using the Sakai-Sugimoto model, showing that indeed the magnetic field has a strong influence over the elliptic flow. Another model was given in \cite{Muller:2013ila}, using a D3/D7 system and including the magnetic field as an excitation over the probe D7-branes. Their results are qualitatively consistent with experimental observations, with the authors stating that this should be regarded as an upper bound for the contribution to $v_2$ that solely a magnetic field could have in a real QGP.

An alternative holographic setup to model the magnetized QGP was introduced in \cite{Avila:2018hsi} as a different 5-dimensional consistent truncation to SUGRA IIB. The constructed family of solutions features the full backreaction of a constant magnetic field and a scalar field dual to an operator of scaling dimension 2. Given that these solutions are part of the same general truncation ansatz studied in \cite{Cvetic:1999xp}, from the gauge theory perspective the magnetic field couples to the $R-$current. A critical intensity for the magnetic field $B_{c}$, above which the system becomes unstable, is induced by the presence of the scalar field. For intensities below $B_{c}$, two branches of solutions exist, with one of them being thermodynamically preferred over the other.

The original motivation for this new model was to find a feasible way of easily adding fundamental degrees of freedom in the probe limit. This latter objective was achieved in \cite{Avila:2019pua,Avila:2020ved}, where it was shown that the interplay between the magnetic and scalar fields leads to a very interesting thermodynamic behavior for the fundamental matter. However, even without considering flavor degrees of freedom the magnetized plasma features rich physics. For instance, we recently showed in \cite{Avila:2021rcu} that for the field content considered there, the photons emitted by the plasma in the adjoint representation are linearly polarized in its state parallel to the reaction plane, the so-called in-plane polarization state, for any non-vanishing magnetic field intensity.

The main objective of this manuscript is to provide additional details of the calculations and extend the analysis in \cite{Avila:2021rcu}. In particular, we relax one of our previously imposed constraints, and present a numerical analysis of the emission rate of photons in both in-plane and out-plane polarization states. Said analysis includes the total energy produced as a function of the magnetic field $B$ and the temperature of the plasma $T$, as well as the magnetic contribution to the elliptic flow $v_{2}$ as a function of the frequency of the emitted photons $\omega$ and the intensity of such field $B$. Regarding the latter, we found that the magnetic field indeed increases $v_{2}$, while for the former we found a very interesting behavior: for any given direction of propagation, the production of photons is increased with respect to the $B=0$ value if the magnetic field intensity lies in the interval $0<B<B_{\vartheta}$, while it decreases for any $B>B_{\vartheta}$. The magnetic field intensity $B_{\vartheta}$ depends on the propagation direction, in a manner such that it tends to zero as the photon momentum is aligned with the magnetic field. This last feature is true regardless of the polarization state.

As mentioned above, the reason why the details omitted in our previous letter \cite{Avila:2021rcu} are rigorously included in the present report is that, as we will see, the fully back-reacted equations involving the in-plane polarization photons are identical to those found there.

The manuscript is organized as follows. In the Sec. \ref{PhotonPlasma} we review the framework to study the emission of photons from the gauge theory perspective. Section \ref{GravityDual} is devoted to the family of gravitational solutions that we use for the holographic description. In Sec. \ref{HolographicPhoton} we present the details of the computations necessary to study the emitted photons using the gauge/gravity correspondence. In Sec. \ref{InPlane}, we focus on the in-plane polarization states, while in Sec. \ref{OutPlane} we deal with the out-plane states. After that we show the results for the total photon production and the elliptic flow in Sec. \ref{Total} and Sec. \ref{EllipticFlow} respectively. We close with a discussion of our results in Sec. \ref{Discussion}. Details about the equations of motion are included in Appendices \ref{AppOut} and \ref{AppIn}.

\section{Photon production in a strongly coupled magnetized plasma}
\label{PhotonPlasma}

The gauge theory we consider is 4-dimensional SYM $\mathcal{N} = 4$ over Minkowski spacetime, with gauge group $SU(N_c)$ at large $N_c$ and 'tHooft coupling $\lambda = g_{\mathrm{YM}}{}^2 N_c$. All the matter fields of this theory are in the adjoint representation of the gauge group. The produced photons are modeled by adding a $U(1)$ kinetic term to the SYM action that couples to the electromagnetic current associated to a $U(1)$ subgroup of the global $SU(4)$ $R$-symmetry group of the theory. Hence, the action adopts the form of a $SU(N_c) \times U(1)$ gauge theory
\begin{equation} 
S=S_{SYM}-\frac{1}{4}\int d^{4}x\left(\mathcal{F}^{2}-4e\mathcal{A}^{\mu}\mathcal{J}^{EM}_{\mu}\right), 
\label{actionSYMU1}
\end{equation}
where $\mathcal{F}=d\mathcal{A}$ is the electromagnetic field, and the electromagnetic current is given by
\begin{equation}
\mathcal{J}^{EM}_{\mu}=\bar{\Psi}\gamma_{\mu}\Psi+\frac{i}{2}\Phi^*(\mathcal{D}_\mu \Phi)-\frac{i}{2} (\mathcal{D}_\mu \Phi^*)^*\Phi.
\label{emcurrent}
\end{equation}
In the previous expression $\Psi$ and $\Phi$ represent the fermionic and scalar fields of SYM $\mathcal{N}=4$, respectively. In this same expression there is an implicit sum over the color indexes, and the derivative operator $\mathcal{D}_{\mu}=D_{\mu}-ieA_{\mu}$ is the covariant derivative of the full $SU(N_{c})\times U(1)$ group.

Given that the electromagnetic coupling $\alpha_{\rm EM} = e^2/4\pi$ is small compared to 'tHooft coupling $\lambda = g_{\rm YM}{}^2 N_c$ (at large $N_c$), even if the two-point correlation function necessary to compute photon production has to be calculated non-perturbatively in the $SU(N_c)$ theory that involves $\lambda$, it is enough to determine it to leading order in $\alpha_{\rm EM}$ and ignore terms of order $\mathcal{O}(\alpha_{\rm EM}^2)$. It is because of this that we can work exclusively in the gravitational dual of the $SU(N_c)$ gauge theory.

If the plasma is in thermal equilibrium at temperature $T$, the rate of emitted photons with wave null 4-vector $k^{\mu}=(k^0,\vec{k})$ and polarization 4-vector $\epsilon^{\nu}_{s}(\vec{k})$ is \cite{CaronHuot:2006te,Mateos:2007yp}
\begin{equation}
\frac{d\Gamma_{s}}{d\vec{k}}=\frac{e^{2}}{(2\pi)^{3}2|\vec{k}|}n_{B}(k^{0})\chi_{s}(k)\bigg|_{k = 0},
\label{diffproduction}
\end{equation}
where
\begin{equation}
\chi_{s}(k)=\epsilon^{\mu}_{s}(\vec{k}) \epsilon^{\nu}_{s}(\vec{k})\chi_{\mu\nu}(k),
\label{SpectralPolarized}
\end{equation}
is the spectral density for the polarization state $\epsilon^{\nu}_{s}(\vec{k})$ and
\begin{equation}
n_{B}(k^{0})=\frac{1}{e^{k^{0}/T}-1},
\label{BoseEinstein}
\end{equation}
is the Bose-Einstein distribution. The spectral density tensor $\chi_{\mu\nu}(k)=-2\mathrm{Im}[G^{R}_{\mu\nu}(k)]$ is given in terms of the retarded two-point correlator of the electromagnetic current \eqref{emcurrent} 
\begin{equation} 
G^{R}_{\mu\nu}(k)=-i\int d^{4}x e^{-ik\cdot x}\Theta(t)\langle\big[\mathcal{J}^{EM}_{\mu}(x),\mathcal{J}^{EM}_{\nu}(0)\big]\rangle,
\label{2pointcorr}
\end{equation}
where the expectation value is taken in the state at temperature $T$. 

\begin{figure}[ht!]
 \centering
 \includegraphics[width=0.5\textwidth]{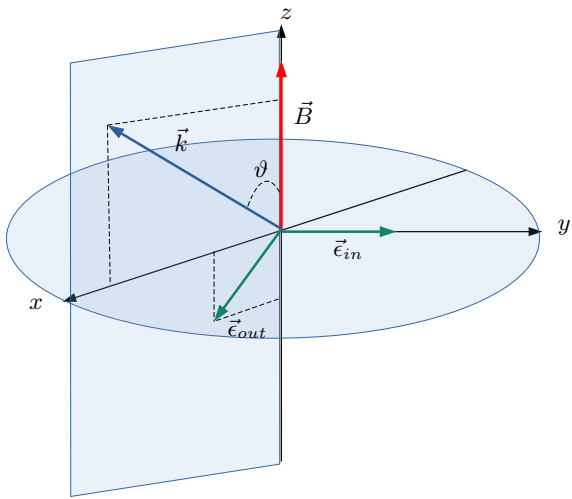}
  \put(-140,190){$z$}
  \put(-35,110){$y$}
  \put(-235,80){$x$}
  \put(-135,150){$\vec{B}$}
  \put(-180,140){$\vec{k}$}
  \put(-160,70){$\vec{\epsilon}_{out}$}
  \put(-120,100){$\vec{\epsilon}_{in}$}
  \put(-150,130){$\vartheta$}
\caption{{\footnotesize Spatial components of the photon momentum $\vec{k}$ and the polarization vectors: $\vec{\epsilon}_{out}$ out-plane and $\vec{\epsilon}_{in}$ in-plane. The magnetic field $\vec{B}$ points perpendicular to the reaction plane, which is depicted as a disk in the $xy$-plane. Given the rotational symmetry around the $z$-direction, without loss of generality we can choose the momentum of the photon parallel to the $xz$-plane.}}
\label{PhotonDiagram}
\end{figure}

The spatial polarization four-vectors $\epsilon_{s}^{\mu}$ are orthogonal to the null wave four-vector with respect to the Minkowski metric, $\epsilon^{i}_{s} k^j \delta_{ij} = 0$ and without loss of generality they can also be chosen to satisfy $\epsilon^{i}_{1} \epsilon^{j}_{2} \delta_{ij} = 0$. We will fix our coordinate system such that the background magnetic field is directed along the $z$-direction. This choice leaves us with a rotational symmetry on the reaction plane ($xy$-plane), allowing us to conveniently set the wave four-vector to lie in the $xz$-plane. Denoting the angle that $\vec{k}$ forms with the background magnetic field by $\vartheta$, the wave and polarization four-vectors take the form 
\begin{eqnarray} 	
&& k^{\mu}=k^{0}(1,\sin\vartheta,0,\cos\vartheta), 
\cr
&& \epsilon_{out}^{\mu}=(0,\cos\vartheta,0,-\sin\vartheta), 
\cr  
&& \epsilon_{in}^{\mu}=(0,0,1,0),
\label{pol_wave_vectors}
\end{eqnarray}
where the notation makes reference to the in-plane and out-plane polarization states. Hence, the spectral density \eqref{SpectralPolarized} for the polarization state $\epsilon_{out}$ is given by
\begin{equation}  
\chi_{out}(k)=\cos^{2}\vartheta \chi_{xx}-2\cos\vartheta\sin\vartheta \chi_{xz}+\sin^{2}\vartheta \chi_{zz}\ ,
\label{SpectralOut}
\end{equation}
while for the $\epsilon_{in}$ state we have
\begin{equation}   
\chi_{in}(k)=\chi_{yy}.
\label{SpectralInn}
\end{equation}
We present a diagram with the described kinematics in FIG. \ref{PhotonDiagram}.

An important quantity which characterizes the anisotropic properties of the medium where the photons are produced is the elliptic flow $v_2$, which gives a measure of the degree of the momentum anisotropy of the particles generated in the collision. Given the total differential rate of emitted photons
\begin{equation}
\frac{d\Gamma}{d\vec{k}}=\frac{d\Gamma_{in}}{d\vec{k}}+\frac{d\Gamma_{out}}{d\vec{k}},
\label{GammaTotal}
\end{equation}
the elliptic flow $v_{2}$ is given by the coefficient of the second harmonic in the Fourier expansion in the azimuthal photon distribution \cite{Lohner:2012ct,PHENIX:2015igl}, which for central rapidity is
\begin{equation}
\frac{d\Gamma}{d\vec{k}}=\Gamma_{0}(1-2v_{2}\cos(2\theta)+2v_{4}\cos(4\theta)+\cdots).
\label{FourierExpansion}
\end{equation}
The minus sign in front of $v_{2}$ comes from the fact that the Fourier expansion is done with respect to $\vartheta$ which is the angle between the photon's wave vector $\vec{k}$ and the magnetic field $\vec{B}$, instead of using $\phi$, the angle between $\vec{k}$ and the reaction plane (where the relation between the two angles is $\vartheta=\frac{\pi}{2} - \phi$). In this way, the elliptic flow can be obtained from $d\Gamma/d\vec{k}$ by projecting it over $\cos(2\theta)$ 
\begin{equation}
v_{2}=-\frac{2}{\pi \Gamma_{0}}\int_{0}^{\frac{\pi}{2}}d\vartheta \cos(2\vartheta)\frac{d\Gamma}{d\vec{k}}=-\frac{\int_{0}^{\frac{\pi}{2}}d\vartheta \cos(2\vartheta)\frac{d\Gamma}{d\vec{k}}}{\int_{0}^{\frac{\pi}{2}}d\vartheta\frac{d\Gamma}{d\vec{k}}}.
\label{v2}
\end{equation}

\section{The gravitational background}
\label{GravityDual}
The holographic model that we consider is a family of solutions to five-dimensional gauged supergravity with its bosonic part of the action given by \cite{Avila:2018hsi}
\begin{equation}
\begin{split}
S=& \frac{1}{16\pi G_{5}}\int d^{5}x \sqrt{-g}\left[R-\frac{1}{2}(\partial\varphi)^{2}\right. \\& \left. +\frac{4}{L^{2}}\left(e^{\frac{2}{\sqrt{6}}\varphi}+2e^{-\frac{1}{\sqrt{6}}\varphi}\right)-e^{-\frac{2}{\sqrt{6}}\varphi}(F)^{2}\right],
\end{split}
\label{Action}
\end{equation}
where the field content consists of the scalar $\varphi$, the Faraday tensor $F$, and the components of the metric $g_{\mu\nu}$. $G_{5}$ is the five-dimensional Newton constant and $L$ is the $AdS_{5}$ radius, which in what follows is set equal to one, $L=1$, and therefore $G_{5}=\frac{\pi}{2N_{c}^{2}}$.

As shown in \cite{Avila:2018hsi}, this family of solutions can be uplifted to ten-dimensional supergravity using the results in \cite{Cvetic:1999xp}, while in five dimensions its elements take the general form
\begin{eqnarray}
&& ds^{2}_{5}=\frac{dr^{2}}{U(r)}-U(r)dt^{2}+V(r)(dx^{2}+dy^{2})+W(r)dz^{2}, 
\cr
&& F=B\,dx\wedge dy,
\cr
&& \varphi=\varphi(r).
\label{Background}
\end{eqnarray}
where $r$ is the $AdS_{5}$ radial coordinate, in terms of which the boundary is located at $r\rightarrow\infty$.  All these backgrounds feature a black hole, with a horizon located at $r=r_{h}$ where the metric function $U(r)$ vanishes. Hence the temperature of each solution is given by
\begin{equation}
T=\frac{3 r_{h}}{2\pi}.
\label{Temperature}
\end{equation}
The magnetic field intensity $B$ coincides with the one in the dual gauge theory given that the metric in \eqref{Background} asymptotes 5-dimensional anti-de Sitter spacetime at the boundary. From the ten-dimensional perspective the Maxwell field is interpreted as an infinitesimal rotation in the compact part of the geometry (see \cite{Cvetic:1999xp,Avila:2020ved} for additional details). Given that the equations of motion coming from \eqref{Action} are highly non-linear, the family of solutions given by \eqref{Background} must be obtained numerically for any non-vanishing intensity of the magnetic field. The general integration procedure is described in detail in \cite{Avila:2018hsi}. It is important to mention that the equations of motion deduced from \eqref{Action} require a non-constant scalar field $\varphi(r)$ for any non-vanishing magnetic field. This means that the gravitational model found in \cite{DHoker:2009mmn} cannot be recovered from ours for $B$ other than zero. For $B=0$, $\varphi=0$ and any $T$  the geometries reduce to the non-compact part of the black D3-brane metric. 

The near boundary behavior of the scalar field is
\begin{equation}
\varphi\rightarrow\frac{1}{r^{2}}\left(\varphi_{0}+\psi_{0}\log{r}\right),
\label{phiasy}
\end{equation}
where the coefficients $\varphi_{0}$ and $\psi_{0}$ can be read from the asymptotics of any specific numerical solution. This behavior implies that $\varphi$ saturates the BF bound \cite{Breitenlohner:1982jf,Bianchi:2001kw}. Hence the scalar field $\varphi$ is dual to a scalar operator $\mathcal{O}_{\varphi}$ of scaling dimension $\Delta=2$. According to the holographic dictionary, $\psi_{0}$ is dual to the source of the operator and $\varphi_{0}$ to its vacuum expectation value $\langle \mathcal{O}_{\varphi}\rangle$ \cite{Bianchi:2001kw}. From the gauge theory perspective, it makes sense to specify the source of the operator and then compute the vacuum expectation value that it generates in response to such source. 

It was shown in \cite{Avila:2018hsi} that for any given $\psi_{0}$ there exists a critical magnetic field intensity $B_{c}$ that the plasma can tolerate, as it becomes unstable for higher values. From the dual gravitational perspective, beyond this critical value $B_{c}$ the geometries develop a naked singularity. Below $B_{c}$ there are two branches of solutions for any fixed $B/T^{2}$ that differ in the value of $\langle \mathcal{O}_{\varphi}\rangle/T^{2}$. In \cite{Avila:2018hsi} it was also shown that one of these branches is thermodynamically preferred over the other. The one with the higher value for $\langle \mathcal{O}_{\varphi}\rangle/T^{2}$ corresponds to a state with negative specific heat, higher free energy and lower entropy than the other, showing that the solutions with smaller $\langle \mathcal{O}_{\varphi}\rangle/T^{2}$ are thermodynamically preferred. Throughout this manuscript we will fix the scalar source to $\psi_{0}=0$ and work exclusively on the thermodynamically favored branch. For this value of the scalar source the maximum magnetic field intensity is given by $B_{c}/T^{2}\simeq 11.24$.

\section{Holographic photon production}
\label{HolographicPhoton}

According to the holographic dictionary, the correlation function \eqref{2pointcorr} can be determined by a perturbative calculation on the gravitational dual \cite{Policastro:2002se,Policastro:2002tn,Son:2002sd,Kovtun:2005ev}. To this end we write
\begin{eqnarray}
&& g_{mn}={g^{BG}}_{mn}+h_{mn},
\cr
&& F=F^{BG}+dA,
\cr
&& \varphi=\varphi^{BG}+\phi,
\label{Perturbations}
\end{eqnarray}
where the superscript $BG$ labels the background fields as given by \eqref{Background} while $h_{mn}$, $A$, and $\phi$ are their first-order perturbations, for which we will solve the field equations at the corresponding order.

The general perturbation of our family of solutions admitted by the five-dimensional theory that results from the dimensional reduction of ten-dimensional supergravity includes a second $U(1)$ field in addition to the one we have considered so far. Nonetheless, the equations of motion decouple in such a manner that the latter $U(1)$ field can consistently be set to zero, which is the path we followed in \cite{Avila:2021rcu} and lead to the results therein.

A full treatment of the perturbations in the truncated theory would demand the inclusion of said additional field, which, unlike the previously considered $U(1)$, couples directly to the scalar field. This construction conducts to a theory with rich physics that we will present shortly \cite{Avila2come}, while in the present work we should limit ourselves to study those perturbations consistent with the variation of the action \eqref{Action} with respect of the fields already included.

In the context that we just have set, the perturbation to the Maxwell field $A$ evaluated at the boundary, which we will denote as $A^{bdry}$, is dual to the source term of the gauge field $\mathcal{A}$ in the dual theory \eqref{actionSYMU1}. In other words, $A^{bdry}$ is dual to the electromagnetic current $\mathcal{J}^{\rm EM}$ in the gauge theory side as long as we work in the $A_{r}=0$ gauge. Moreover, given that the boundary is perpendicular to $r$, we will also impose the gauge $h_{mr}=0$ for the metric perturbations. The previous gauge choices are consistent with the fact that these components are not dual to a source on the gauge theory.

The procedure to compute the production of photons is as follows: The equations of motion from \eqref{Action} are numerically solved at first order in the perturbations $h_{mn}$, $A$, and $\phi$ in \eqref{Perturbations}. Since we are interested in the retarded correlator, we consider only the corresponding ingoing and regular solutions obtained by the Frobenius method applied near the horizon, discarding the outgoing modes. Next, in order to obtain the correlator \eqref{2pointcorr} it is necessary to evaluate the action \eqref{Action} on the perturbed solutions and take the second variation with respect to $A^{bdry}$. Since all the perturbations are coupled by the equations of motion, we apply the method to compute correlators for mixing operators developed in \cite{Amado:2009ts,Kaminski:2009dh}.

First, the near boundary behavior of the equations reveal that the asymptotic solutions for the scalar field and metric perturbations are
\begin{equation}
\begin{split}
& h_{mn}(r)=r^{2}\tilde{h}_{mn}(r), \\
& \phi(r)=\frac{1}{r^{2}}\log(r)\tilde{\phi}(r)
\end{split}
\label{rescaling}
\end{equation}
where $\tilde{h}_{mn}(r)$ and $\tilde{\phi}(r)$ tend to constant values as $r\rightarrow\infty$. In order to follow the procedure outlined in \cite{Arciniega:2013dqa,Amado:2009ts,Kaminski:2009dh} we must solve for the normalized fields $\tilde{h}_{mn}(r)$ and $\tilde{\phi}(r)$. To keep a clean notation, we will refer to these normalized fields as $h_{mn}(r)$ and $\phi(r)$ hoping that no confusion will arise.

We then look for all the linearly independent non-outgoing solutions such that their behavior near the boundary is 
\begin{equation}
\lim_{r\rightarrow\infty}\begin{pmatrix}
A^{(1)} \\ \Phi^{(1)}  
\end{pmatrix}
=\begin{pmatrix}
1 \\ 0
\end{pmatrix}, \qquad 
\lim_{r\rightarrow\infty}\begin{pmatrix}
A^{(2)} \\ \Phi^{(2)}  
\end{pmatrix}
=\begin{pmatrix}
0 \\ 1
\end{pmatrix},
\label{SolLI}
\end{equation}
where $A$ denotes any of the components of the gauge field perturbation and $\Phi$ denotes any of the fields coupled to $A$. Given that we are solving the equations for the perturbations at linear order, the most general solution can be written as a linear combination of \eqref{SolLI}
\begin{equation}
\begin{pmatrix}
A \\ \Phi
\end{pmatrix}=
A^{bdry}\begin{pmatrix}
A^{(1)} \\ \Phi^{(1)}  
\end{pmatrix}+
\Phi^{bdry}\begin{pmatrix}
A^{(2)} \\ \Phi^{(2)}  
\end{pmatrix}\ .
\label{GeneralSol}
\end{equation}
Note that in this way the $A^{bdry}$ dependence has been made explicit, making it simple to take variations with respect to this quantity. We can see from \eqref{GeneralSol} that the values the fields take at the boundary are independent of each other
\begin{equation}
\frac{\delta A|_{bdry}}{\delta A^{bdry}}=1, \qquad \frac{\delta \Phi|_{bdry}}{\delta A^{bdry}}=0,
\end{equation}
while this is not the case for the derivatives
\begin{equation}
\frac{\delta A'|_{bdry}}{\delta A^{bdry}}=A'^{(1)}|_{bdry}, \qquad \frac{\delta \Phi'|_{bdry}}{\delta A^{bdry}}=\Phi'^{(1)}|_{bdry},
\end{equation}
where a prime denotes the derivative with respect to $r$. We will use this generic fact to compute the retarded Green function from the action \eqref{Action}. In Sec. \ref{InPlane} we will show how to explicitly construct the solutions in \eqref{SolLI}.

After evaluating \eqref{Action} in the perturbed solutions, we look for the second derivative terms and integrate them by parts to obtain a boundary term, keeping the calculations at second order in the perturbations. Schematically we have
\begin{equation} 
\label{scheaction}
S^{bdry}\propto\int d^{4}x(\mathcal{O}(A A')+\mathcal{O}(\phi\phi')+\mathcal{O}(h^{2})+\mathcal{O}(hh')),
\end{equation} 
where the limit $r\rightarrow\infty$ is implicit and the zero, first and greater than second order terms have not been written, as they are not relevant to our computation. From the previous discussion we see that the $\mathcal{O}(\phi\phi')$, $\mathcal{O}(h^{2})$ and $\mathcal{O}(hh')$ terms do not contribute to the Green function, as they vanish when taking the second variation with respect to $A^{bdry}$, thus the only relevant part of the boundary action is
\begin{equation} 
\begin{split}
S^{bdry}=&-\frac{1}{8\pi G_5}\int d^4 x\,U V X^{-2}\sqrt{W}\left(-\frac{A_{t}A'_{t}}{U}+\frac{A_{x}A'_{x}}{V}\right.\\&\left.+\frac{A_{y}A'_{y}}{V}+\frac{A_{z}A'_{z}}{W}\right),
\end{split}
\label{BdryAction}
\end{equation}
where $X = e^{\frac{1}{\sqrt{6}}\varphi}$. The only other terms that could have contributed are of the form $\mathcal{O}(h'h')$, $\mathcal{O}(\phi'\phi')$, $\mathcal{O}(\phi'h')$, $\mathcal{O}(A h')$, $\mathcal{O}(A \phi')$, $\mathcal{O}(A'h')$ and $\mathcal{O}(A'\phi')$, but none of them appear in the action.

We need to solve the field equations to first order in the perturbations and find the non-outgoing solutions \eqref{SolLI}. To this end, we note that even if the dual gauge theory is anisotropic because of the presence of the magnetic field, it is still invariant under translations, hence we can take the Fourier decomposition of all the fields
\begin{equation}
\begin{split}
&\Phi(r,x^{\mu})=\int \frac{d^{4}k}{(2\pi)^{4}}e^{-i k_{\mu}x^{\mu}}\Phi(r,k^{\mu}), \\& k^{\mu}=k^{0}(1,\sin\vartheta,0,\cos\vartheta),
\end{split}
\label{Fourier}
\end{equation}
where $\Phi$ denotes any of the fields $h_{mn}$, $A_{m}$ or $\phi$. This results in 21 ordinary differential equations that decouple in two independent groups, corresponding to the independent polarization states. It can be verified that the equations involving the in-plane polarization presented here in Appendix \ref{AppIn}, are identical to those that arise in the restricted setting considered in \cite{Avila:2021rcu}.


\section{In-plane polarization state}
\label{InPlane}
In order to compute the production of photons in the polarization state $\epsilon_{in}$ we need to solve thirteen differential equations for nine components of the fields: $A_{y}$, $\phi$, $h_{tx}$, $h_{tz}$, $h_{xz}$, $h_{tt}$, $h_{xx}$, $h_{yy}$ y $h_{zz}$. Even when the number of equations surpasses the number of variables, the system can be solved consistently. Of the thirteen equations, four are constraints that once imposed at a given $r$, they will be satisfied at any other radial position. This leaves nine second order differential equations, thus once the constraints have been taken into account, we have fourteen linearly independent solutions to the system. We present the equations explicitly in the Appendix \ref{AppIn}.

The system does not have an analytical solution for any non-vanishing magnetic field, and as a consequence we need to resort to numerical methods to obtain a solution. The first step is to solve the equations near the horizon, using a Frobenius expansion around $r_{h}$ of the form 
\begin{equation}
\Phi(r)=(r-r_{h})^{\alpha}\sum_{j=0}^{\infty}\Phi_{j}(r-r_{h})^{j},
\label{Frobenius}
\end{equation}
where $\Phi$ denotes any of the fields. This procedure gives three independent solutions for $\alpha=0$, five ingoing solutions with $\alpha=-ik_{0}/6r_{h}$, five outgoing solutions with $\alpha=ik_{0}/6r_{h}$ and one with $\alpha=1/2$. As a consequence, the space of non-outgoing solutions is 9-dimensional. In what follows, we will denote an arbitrary element of this space as
\begin{equation}
\text{Sol}=\begin{pmatrix}
A_{y} \\ \phi \\ h_{tx} \\ h_{tz} \\ h_{xz} \\ h_{tt} \\ h_{xx} \\ h_{yy} \\ h_{zz}
\end{pmatrix}.
\end{equation}

Next we look for the nine linearly independent non-outgoing solutions \eqref{SolLI} such that its near boundary behavior is
\begin{equation}
\lim_{r\rightarrow\infty}\text{Sol}^{(1)}
=\begin{pmatrix}
1 \\ 0 \\ 0 \\ 0 \\ 0 \\ 0 \\ 0 \\ 0 \\ 0
\end{pmatrix}, \cdots
,\lim_{r\rightarrow\infty}\text{Sol}^{(9)}
=\begin{pmatrix}
0 \\ 0 \\ 0 \\ 0 \\ 0 \\ 0 \\ 0 \\ 0 \\ 1
\end{pmatrix},
\label{SolLIv2}
\end{equation}
hence an arbitrary non-outgoing solution can be written as 
\begin{equation}
\text{Sol}=A_{y}^{bdry}\text{Sol}^{(1)}+\phi^{bdry}\text{Sol}^{(2)}+\ldots+h_{zz}^{bdry}\text{Sol}^{(9)}.
\label{GeneralSolv2}
\end{equation}
From this expression we can compute the variation with respect to $A_{y}^{bdry}$, which gives
\begin{equation}
\frac{\delta \text{Sol}|_{bdry}}{\delta A_{y}^{bdry}}=\text{Sol}^{(1)}|_{bdry}, \qquad \frac{\delta \text{Sol}'|_{bdry}}{\delta A_{y}^{bdry}}=\text{Sol}'^{(1)}|_{bdry}.
\label{Variation}
\end{equation}

We can substitute \eqref{GeneralSolv2} on the action \eqref{BdryAction} and take its second variation with respect to $A^{bdry}$ using \eqref{Variation}. The final result is that the correlator function is
\begin{equation}
G^{R}_{yy}=-\frac{1}{4\pi G_{5}}\left(UX^{-2}\sqrt{W}{A'_{y}}^{(1)}\right) \bigg|_{bdry}.
\label{GRyy}
\end{equation}
It is clear then that we need to compute the solution $\text{Sol}^{(1)}$ explicitly. In order to do this we look for nine non-outgoing solutions with an arbitrary behavior near the boundary, and use them as columns of the matrix
\begin{equation}
\mathcal{M}=(\text{Sol}^{(a1)} \, \text{Sol}^{(a2)} \, \ldots \, \text{Sol}^{(a9)}). 
\label{MatrixM}
\end{equation}
We can use this matrix to invert \eqref{GeneralSolv2} and obtain
\begin{equation}
(\text{Sol}^{(1)} \, \text{Sol}^{(2)} \, \ldots \, \text{Sol}^{(9)})=\mathcal{M}(\mathcal{M}^{-1}|_{bdry}),
\end{equation}
from where $\text{Sol}^{(1)}$ can be read immediately.

The arbitrary solutions that constitute the matrix \eqref{MatrixM} can be constructed by numerically solving the equations of motion. In practice this procedure consists in solving around the horizon using the expansions \eqref{Frobenius}, choosing the values for $\alpha$ that exclude the outgoing solutions. The result of this is used as the initial conditions for the numerical integration from $r=r_{h}+\epsilon$, with $\epsilon\ll r_{h}$, to the boundary at $r\rightarrow\infty$.  

\begin{figure}
\begin{center}
\includegraphics[width=0.4\textwidth]{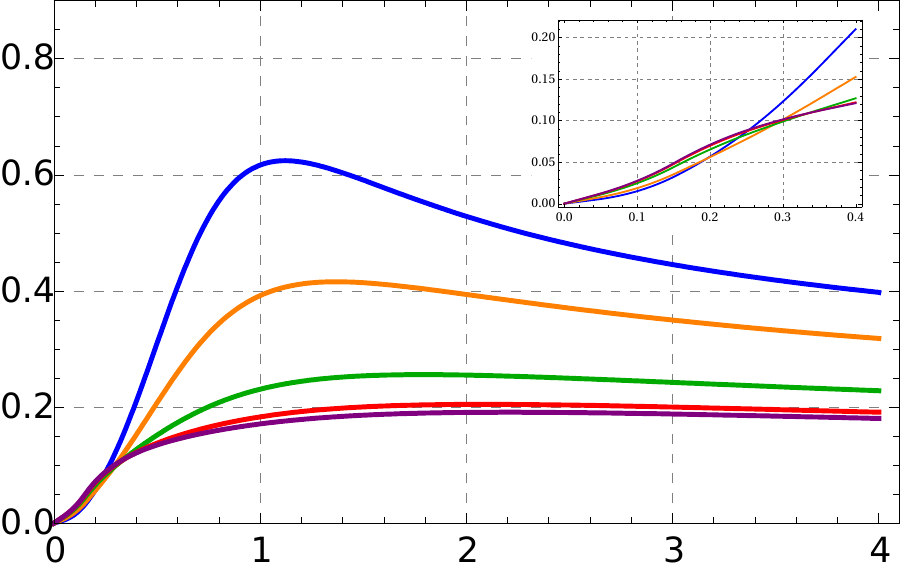} 
   \put(0,-10){$\omega$}
   \put(-225,130){$\frac{\chi_{in}}{2 N_{c}^{2}\omega T^{2}}$}
\end{center}
\caption{\small Spectral function $\chi_{in}$ for the in-plane polarization state in terms of the photon frequency $\omega=k_{0}/2\pi T$ for fixed magnetic field $B/T^{2}=11.24$. The blue, orange, green, red and purple curves (from top to bottom on the right side of the graph) correspond to $\vartheta=\{\pi/2, \pi/4, \pi/8, \pi/16, \pi/32\}$ respectively.}
\label{Chi_bmax}
\end{figure}

\begin{figure}
\begin{center}
\includegraphics[width=0.4\textwidth]{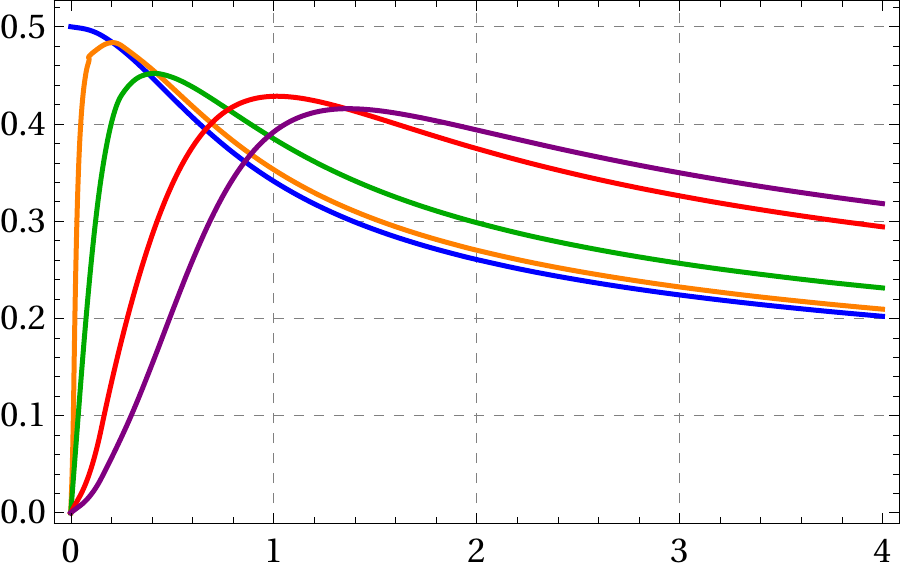} 
   \put(0,-10){$\omega$}
   \put(-225,135){$\frac{\chi_{in}}{2 N_{c}^{2}\omega T^{2}}$}
\end{center}
\caption{\small Spectral function $\chi_{in}$ for the in-plane polarization state in terms of the photon frequency $\omega=k_{0}/2\pi T$ for the propagation direction given by $\vartheta=\pi/4$. The blue, orange, green, red and purple curves (from top to bottom on the left side of the graph) correspond to $B/T^{2}=\{0,2.01,4.22,8.73,11.24\}$ respectively.}
\label{Chi_Pi4}
\end{figure}

With this procedure it is possible to compute the spectral density for the in-plane polarization state \eqref{SpectralInn} as a function of the dimensionless frequency 
\begin{equation}
\omega=\frac{k_{0}}{2\pi T},
\end{equation}
for different values of $\vartheta$ and $B/T^{2}$. In FIG. \ref{Chi_bmax} we show the results for $B/T^{2}=11.24$ and different values of $\vartheta$, each one represented by a different color. We see that for high frequencies the spectral density $\chi_{in}$ decreases as the value of $\vartheta$ diminishes, while for small values of $\omega$ it displays the opposite behavior. In other words, for high values of the frequency, $\chi_{in}$ decreases as the direction of propagation is aligned with the magnetic field, while for small $\omega$ the spectral function decreases as the direction of propagation aligns with the reaction plane. Although we only show here the results for this specific magnetic field intensity, we verified that the same qualitative behavior is shared for any $B/T^{2}\neq 0$.

In FIG. \ref{Chi_Pi4} we show the results for $\vartheta=\pi/4$ and several values of the magnetic field, each one represented by a different colored curve. We see that for small frequencies, $\chi_{in}$ decreases as the magnetic field intensity increases, while for high frequencies this behavior is reversed. Rephrasing the result, for small frequencies the spectral function displays inverse magnetic catalysis (IMC), whereas for high frequencies we observe magnetic catalysis (MC). We present here the results only for $\vartheta=\pi/4$. However, we verified that the same qualitative behavior is reproduced for any $\vartheta$. 

Having computed the spectral function for the in-plane polarization state, we can now proceed to calculate the corresponding differential emission of photons. Given that we have rotational invariance on the reaction plane, we can integrate over this direction in \eqref{diffproduction} to obtain
\begin{equation}
\frac{G_{5}}{2\alpha_{EM}T^{3}}\frac{d\Gamma_{in}}{dk_{0}d\cos\vartheta}=\frac{G_{5}\omega}{2T^{2}}n_{B}(k_{0})\chi_{in}.
\label{GammaInt1}
\end{equation}
We show the results for this quantity as a function of the photon frequency and different values for the magnetic field intensity and propagation directions.

\begin{figure}
\begin{center}
\includegraphics[width=0.4\textwidth]{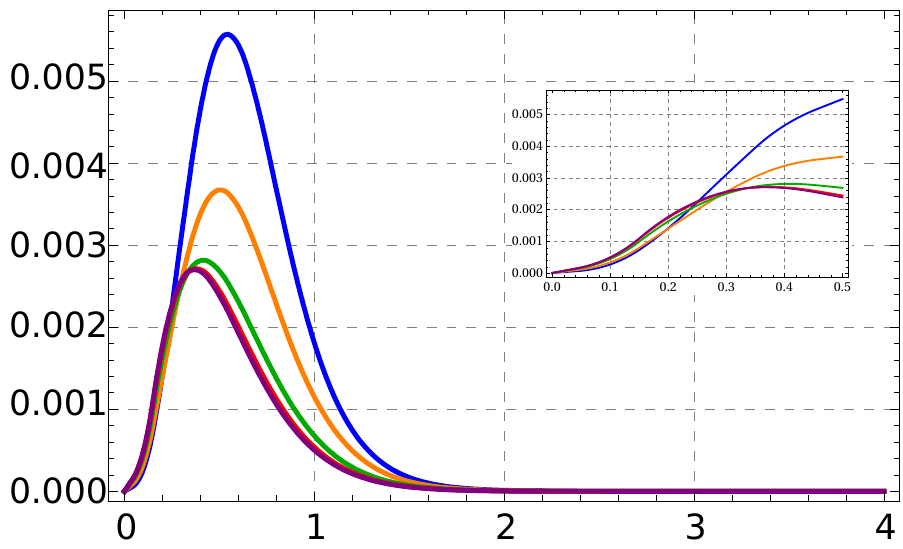} 
   \put(0,-10){$\omega$}
   \put(-220,130){$\frac{1}{8 N_{c}^{2}\alpha_{EM}T^{3}}\frac{d\Gamma_{in}}{dk_{0}d\cos\vartheta}$}
\end{center}
\caption{\small Differential photon production $\frac{d\Gamma_{in}}{dk_{0}d\cos\vartheta}$ for the in-plane polarization state as a function of the photon frequency $\omega=k_{0}/2\pi T$ for fixed magnetic field at
$B/T^{2}=11.24$. The blue, orange, green, red and purple curves (from top to bottom in the middle) correspond to $\vartheta=\{\pi/2, \pi/4, \pi/8, \pi/16, \pi/32\}$ respectively.}
\label{Gamma_bmax}
\end{figure}

\begin{figure}
\begin{center}
\includegraphics[width=0.4\textwidth]{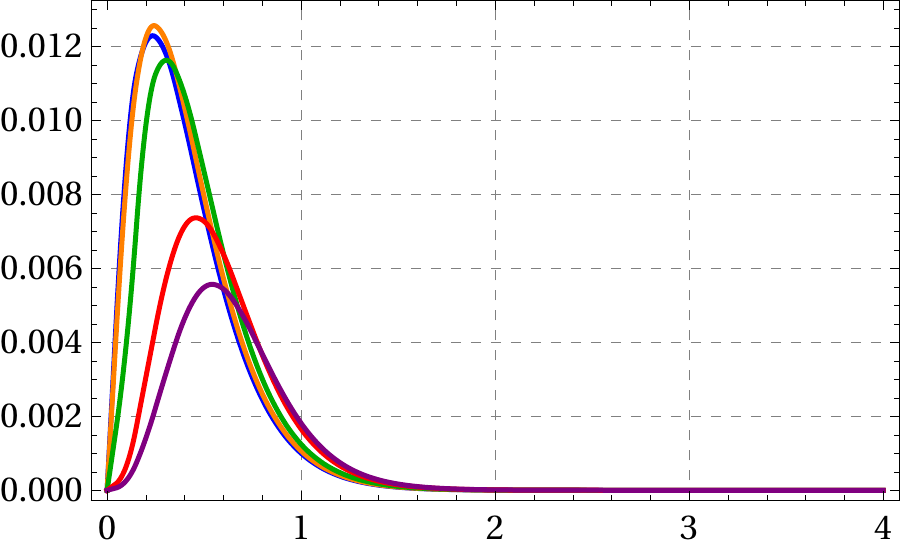} 
   \put(0,-10){$\omega$}
   \put(-220,130){$\frac{1}{8 N_{c}^{2}\alpha_{EM}T^{3}}\frac{d\Gamma_{in}}{dk_{0}d\cos\vartheta}$}
\end{center}
\caption{\small Differential photon production $\frac{d\Gamma_{in}}{dk_{0}d\cos\vartheta}$ for the in-plane polarization state as a function of the photon frequency $\omega=k_{0}/2\pi T$ for fixed propagation direction at $\vartheta=\pi/2$. The blue, orange, green, red and purple curves (from top to bottom on the left side of the graph) correspond to $B/T^{2}=\{0,2.01,4.22,8.73,11.24\}$ respectively.}
\label{Gamma_Pi2}
\end{figure}

In FIG. \ref{Gamma_bmax} we show the differential photon production for the in-plane polarization state as a function of $\omega$ for fixed magnetic field at $B/T^{2}=11.24$. The different curves correspond to different values of the propagation angle $\vartheta$. We can see how for small frequencies $0 < \omega <0.3$, the photon production is increased as the propagation direction is aligned with the magnetic field. However, for higher frequencies this behavior is reversed, as the production of photons is increased as the direction of propagation aligns with the reaction plane, reaching its maximum precisely when $\vartheta=\pi/2$. Even if we only show here the results for $B/T^{2}=11.24$, we verified that the same qualitative behavior is shared by any other non-zero magnetic field. Next, in FIG. \ref{Gamma_Pi2} we show the results for $d\Gamma_{in}/(dk_{0}d\cos\vartheta)$ at fixed $\vartheta=\pi/2$ and different magnetic field intensities. In can be seen how increasing the magnetic field decreases the production of photons for small frequencies, that is, we have IMC. However, for high frequencies the opposite behavior is displayed and the photon production is increased with the magnetic field, hence the phenomenom of MC is present.

\begin{figure}
\begin{center}
\includegraphics[width=0.4\textwidth]{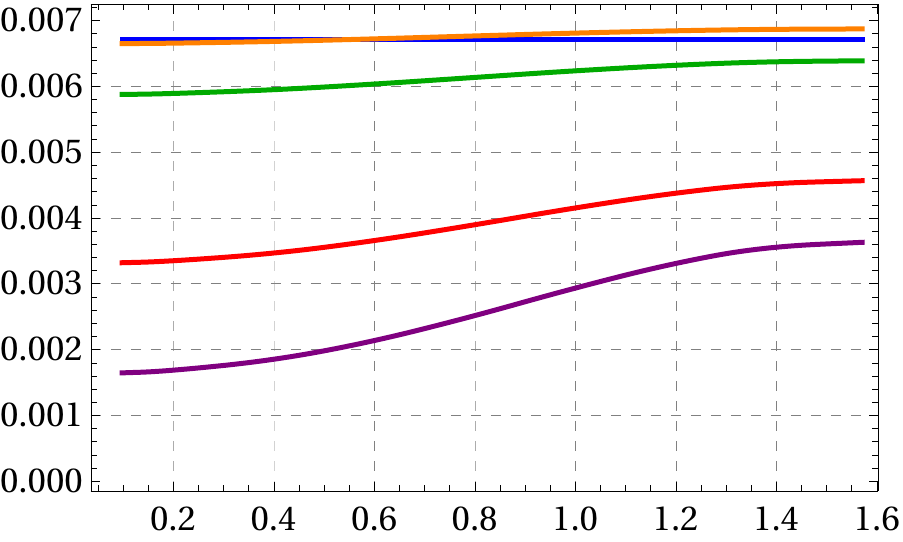} 
   \put(0,-10){$\vartheta$}
   \put(-220,130){$\frac{1}{8 N_{c}^{2}\alpha_{EM}T^{4}}\frac{d\Gamma_{in}}{d\cos\vartheta}$}
\end{center}
\caption{\small Differential photon production $\frac{d\Gamma_{in}}{d\cos\vartheta}$ for the in-plane polarization state as a function of $\vartheta$. The blue, orange, green, red and purple (from top to bottom on the left side of the graph) correspond to $B/T^{2}=\{0,2.01,4.22,8.73,11.24\}$ respectively.}
\label{Gamma_Theta}
\end{figure}

So far we have examined the different quantities as a function of the photon frequency. Nonetheless, it is possible to integrate $\omega$ in \eqref{GammaInt1} to obtain the differential photon production at a given direction and fixed magnetic field. In FIG. \ref{Gamma_Theta} we show this quantity for the in-plane polarization state as a function of the propagation angle $\vartheta$ for different magnetic field intensities. For instance, for the values $B/T^{2}=\{4.22, 8.73, 11.24\}$ we have the same qualitative behavior, that is, the production of photons increases as the momentum is aligned with the reaction plane. Also, for these magnetic field intensities the production is always less when compared to the $B=0$ case for any $\vartheta$. However, for $B/T^{2}=2.01$ this last behavior changes. While it is still true that the production increases as $\vartheta\rightarrow\pi/2$, for $\vartheta>\pi/4$ more photons are produced when compared to the $B=0$ case.

\begin{figure}
\begin{center}
\includegraphics[width=0.4\textwidth]{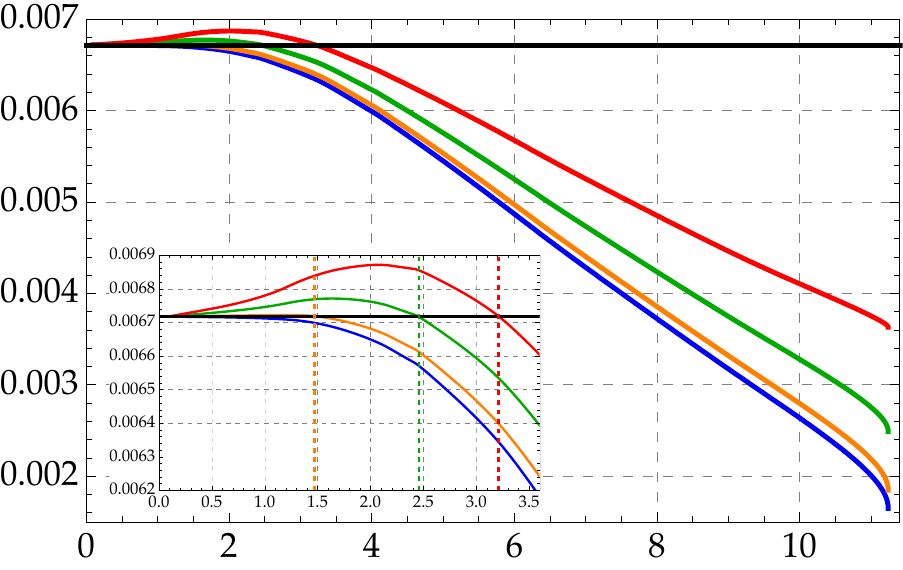} 
   \put(0,-10){$\frac{B}{T^{2}}$}
   \put(-220,140){$\frac{1}{8 N_{c}^{2}\alpha_{EM}T^{4}}\frac{d\Gamma_{in}}{d\cos\vartheta}$}
\end{center}
\caption{\small Differential photon production $\frac{d\Gamma_{in}}{dk_{0}d\cos\vartheta}$ for the in-plane polarization state as a function of $B/T^{2}$. The blue, orange, green and red curves (from bottom to top on the right side of the graph) correspond to $\vartheta=\{\pi/32, \pi/8, \pi/4, \pi/2\}$, respectively. The horizontal black line is a reference for the total photon production for $B/T^{2}=0$. The vertical dashed lines correspond to the $B^{in}_{\vartheta}$ magnetic field intensities below which the production of photons is enhanced with respect to the $B=0$ case. For the propagation angles considered we have $B^{in}_{\pi/2}/T^{2}=3.21$, $B^{in}_{\pi/4}/T^{2}=2.46$ and $B^{in}_{\pi/8}/T^{2}=1.47$.}
\label{Gamma_b}
\end{figure}

The previously described phenomenon can be better appreciated in FIG. \ref{Gamma_b}, where we show the differential photon production for the in-plane polarization state as a function of the magnetic field for different directions of propagation. It can be seen that for any fixed $\vartheta$, there exists a magnetic field intensity $B^{in}_{\vartheta}$ such as if $B>B^{in}_{\vartheta}$ less photons are produced when compared to the $B=0$ case. However, if $0<B<B^{in}_{\vartheta}$ the magnetic field enhances the production of photons. Up to our numerical precision, we were able to corroborate that $B^{in}_{\vartheta}\rightarrow 0$ as $\vartheta\rightarrow 0$, that is, as the photon momentum is aligned with the magnetic field.

\section{Out-plane polarization state}
\label{OutPlane}

In order to compute the production of photons in the out-plane polarization state $\epsilon_{out}$, we need to solve eight equations for the fields $A_{t}$, $A_{x}$, $A_{z}$, $h_{ty}$, $h_{xy}$ and $h_{yz}$. Once again, the number of equations surpasses the number of fields. When we previously studied these gravitational perturbation equations in \cite{Avila:2021rcu}, we imposed the constraint $F\wedge F=0$ that resulted from setting the second $U(1)$ field to zero, showing that the emitted photons were linearly polarized, as no photons could be in the polarization state $\epsilon_{out}$. The equations considered here, listed explicitly in Appendix \ref{AppOut}, do not have such constraint.

In the system of eight equations, two are constraints that once imposed at a given $r$ will be satisfied at all radial positions, while the remaining six are of second order. Hence, once the constraints have been taken into account we are left with ten linearly independent solutions to the system. The Frobenius expansion \eqref{Frobenius} near the horizon gives two independent solutions for $\alpha=0$, four ingoing solutions with $\alpha=-ik_{0}/6r_{h}$ and four outgoing solutions with $\alpha=ik_{0}/6r_{h}$, meaning that we are left with a 6-dimensional space of non-outgoing solutions.

We proceed as in Sec. \ref{InPlane} and denote an arbitrary solution as
\begin{equation}
\text{Sol}=\begin{pmatrix}
A_{t} \\ A_{x} \\ A_{z} \\ h_{ty} \\ h_{xy} \\ h_{yz}
\end{pmatrix} ,
\end{equation}
and look for the basis of non-outgoing solutions \eqref{SolLI} such that its near-boundary behavior is
\begin{equation}
\lim_{r\rightarrow\infty}\text{Sol}^{(1)}
=\begin{pmatrix}
1 \\ 0 \\ 0 \\ 0 \\ 0 \\ 0
\end{pmatrix}, \cdots
,\lim_{r\rightarrow\infty}\text{Sol}^{(6)}
=\begin{pmatrix}
0 \\ 0 \\ 0 \\ 0 \\ 0 \\ 1
\end{pmatrix}.
\label{SolLIOut}
\end{equation}
We can write an arbitrary non-outgoing solution in terms of \eqref{SolLIOut} as
\begin{equation}
\text{Sol}=A_{t}^{bdry}\text{Sol}^{(1)}+A_{x}^{bdry}\text{Sol}^{(2)}+\ldots+h_{yz}^{bdry}\text{Sol}^{(6)}.
\label{GeneralSolOut}
\end{equation}
The solution expressed in this form allows us to easily compute the variation with respect to $A_{x}^{bdry}$ or $A_{z}^{bdry}$, obtaining the following relations
\begin{equation}
\begin{split}
& \frac{\delta \text{Sol}|_{bdry}}{\delta A_{x}^{bdry}}=\text{Sol}^{(2)}|_{bdry}, \qquad \frac{\delta \text{Sol}'|_{bdry}}{\delta A_{x}^{bdry}}=\text{Sol}'^{(2)}|_{bdry},\\
& \frac{\delta \text{Sol}|_{bdry}}{\delta A_{z}^{bdry}}=\text{Sol}^{(3)}|_{bdry}, \qquad \frac{\delta \text{Sol}'|_{bdry}}{\delta A_{z}^{bdry}}=\text{Sol}'^{(3)}|_{bdry}.
\end{split}
\label{VariationOut}
\end{equation}

After substitution of \eqref{GeneralSolOut} in the action \eqref{BdryAction}, the Green functions coming from the variations with respect to $A_{x}^{bdry}$ and $A_{z}^{bdry}$ are given by
\begin{equation}
\begin{split}
& G^{R}_{xx}=-\frac{1}{4\pi G_{5}}\left(U\sqrt{W}X^{-2}{A'_{x}}^{(2)}\right) \bigg|_{bdry}, \\
& G^{R}_{zz}=-\frac{1}{4\pi G_{5}}\left(\frac{U VX^{-2}}{\sqrt{W}}{A'_{z}}^{(3)}\right) \bigg|_{bdry}, \\
& G^{R}_{xz}=-\frac{1}{8\pi G_{5}}\left(U V \sqrt{W}X^{-2}\left(\frac{{A'_{x}}^{(3)}}{V}+\frac{{A'_{z}}^{(2)}}{W}\right)\right) \bigg|_{bdry}. 
\end{split}
\label{GROut}
\end{equation}
In order to compute the solutions $\text{Sol}^{(2)}$ and $\text{Sol}^{(3)}$ explicitly, we proceed as before and look for six non-outgoing solutions with an arbitrary behavior near the boundary. Using these solutions as columns of a matrix analogous to \eqref{MatrixM}, we can invert \eqref{GeneralSolOut} and read $\text{Sol}^{(2)}$ and $\text{Sol}^{(3)}$ from it. The six arbitrary non-outgoing solutions can be obtained numerically solving the equations of motion following the same procedure described in Sec. \ref{InPlane}. To compute the spectral density for the out-plane polarization state \eqref{SpectralOut} we require only the Green functions listed in \eqref{GROut}, so we have all the elements necessary to plot $\chi_{out}$ for different values of $\omega$, $\vartheta$ and $B/T^{2}$.

\begin{figure}
\begin{center}
\includegraphics[width=0.4\textwidth]{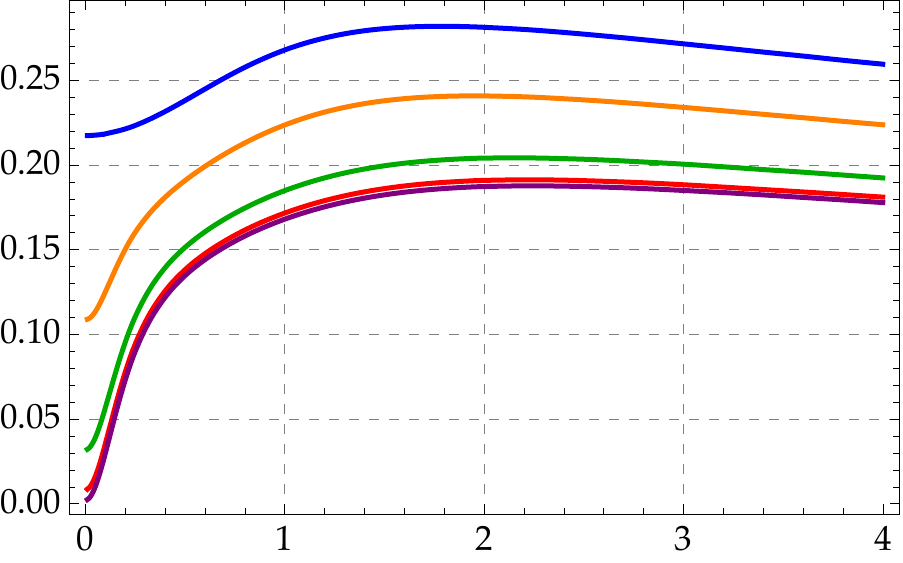} 
   \put(0,-10){$\omega$}
   \put(-225,130){$\frac{\chi_{out}}{2 N_{c}^{2}\omega T^{2}}$}
\end{center}
\caption{\small Spectral function $\chi_{out}$ for the out-plane polarization state in terms of the photon frequency $\omega=k_{0}/2\pi T$ for fixed magnetic field $B/T^{2}=11.24$. The blue, orange, green, red and purple curves (from top to bottom) correspond to $\vartheta=\{\pi/2, \pi/4, \pi/8, \pi/16, \pi/32\}$ respectively.}
\label{Chi_Out_bmax}
\end{figure}

\begin{figure}
\begin{center}
\includegraphics[width=0.4\textwidth]{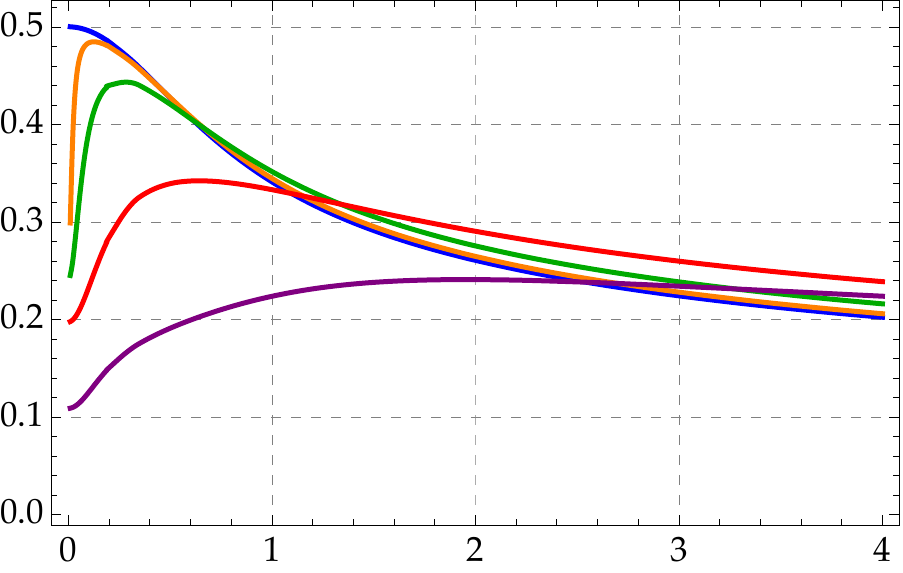} 
   \put(0,-10){$\omega$}
   \put(-225,135){$\frac{\chi_{out}}{2 N_{c}^{2}\omega T^{2}}$}
\end{center}
\caption{\small Spectral function $\chi_{out}$ for the out-plane polarization state in terms of the photon frequency $\omega=k_{0}/2\pi T$ for the propagation direction given by $\vartheta=\pi/4$. The blue, orange, green, red and purple curves (from top to bottom on the left side of the graph) correspond to $B/T^{2}=\{0,2.01,4.22,8.73,11.24\}$ respectively.}
\label{Chi_Out_Pi4}
\end{figure}

In FIG. \ref{Chi_Out_bmax} we show $\chi_{out}$ as a function of the dimensionless photon frequency $\omega$ for fixed magnetic field $B/T^{2}=11.24$ and various values for the propagation angle $\vartheta$. It can be seen that for all $\omega$ the spectral function decreases as the direction of propagation of the photons aligns with the magnetic field. This behavior is in stark contrast with the in-plane polarization state, for which the dependence on the frequency is monotonous. Here we show only the results for $B/T^{2}=11.24$, however, we explicitly verified that this qualitative behavior is reproduced for many intensities of the magnetic field, allowing us to infer that this characteristic feature is valid for any value $B/T^2$ below $B_c$. Meanwhile, in FIG. \ref{Chi_Out_Pi4} we display the results for $\vartheta=\pi/4$ and several values of the magnetic field, each one represented by a different colored curve. We can see that for small frequencies, $\chi_{out}$ decreases as the magnetic field intensity increases, while for high frequencies we have the opposite behavior. In other words, for small frequencies the spectral function for the out-plane polarization state displays IMC, whereas for high frequencies we observe MC, the same as it was for the in-plane state. We present here the results only for $\vartheta=\pi/4$, however, we verified that the same qualitative behavior is reproduced for many directions $\vartheta$, indicating that this qualitative behavior is common to all propagation directions.

\begin{figure}
\begin{center}
\includegraphics[width=0.4\textwidth]{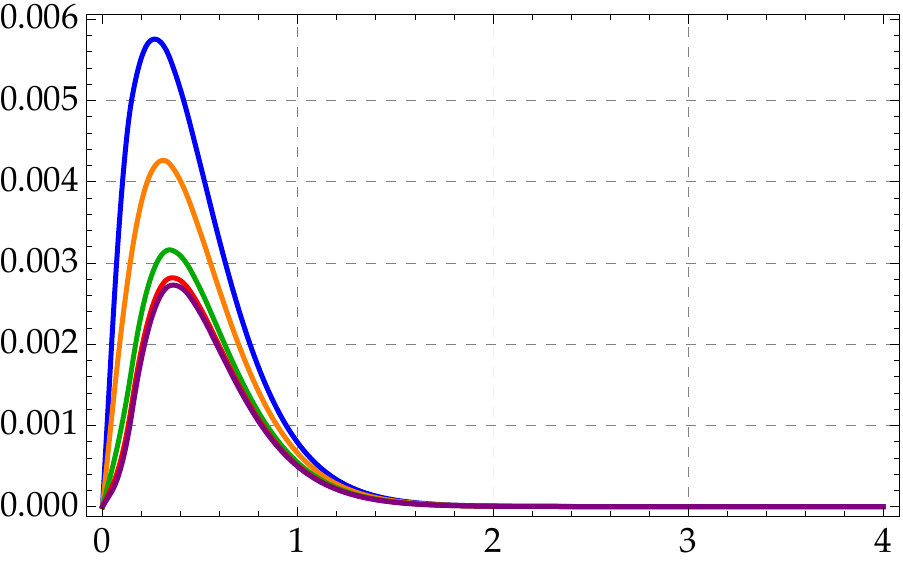} 
   \put(0,-10){$\omega$}
   \put(-220,140){$\frac{1}{8 N_{c}^{2}\alpha_{EM}T^{3}}\frac{d\Gamma_{out}}{dk_{0}d\cos\vartheta}$}
\end{center}
\caption{\small Differential photon production $\frac{d\Gamma_{out}}{dk_{0}d\cos\vartheta}$ for the out-plane polarization state as a function of the photon frequency $\omega=k_{0}/2\pi T$ for fixed magnetic field at
$B/T^{2}=11.24$. The blue, orange, green, red and purple curves (from top to bottom) correspond to $\vartheta=\{\pi/2, \pi/4, \pi/8, \pi/16, \pi/32\}$ respectively.}
\label{Gamma_Out_bmax}
\end{figure}

\begin{figure}
\begin{center}
\includegraphics[width=0.4\textwidth]{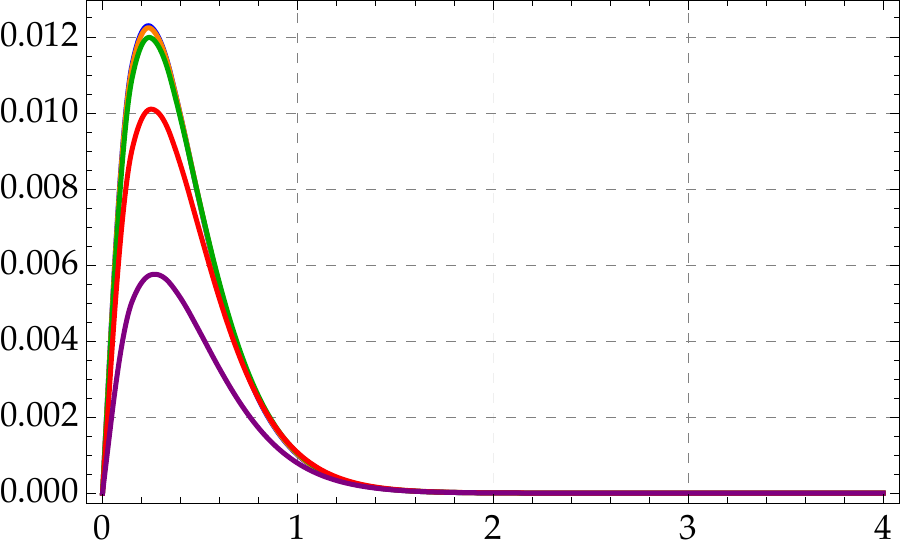} 
   \put(0,-10){$\omega$}
   \put(-220,130){$\frac{1}{8 N_{c}^{2}\alpha_{EM}T^{3}}\frac{d\Gamma_{out}}{dk_{0}d\cos\vartheta}$}
\end{center}
\caption{\small Differential photon production $\frac{d\Gamma_{out}}{dk_{0}d\cos\vartheta}$ for the out-plane polarization state as a function of the photon frequency $\omega=k_{0}/2\pi T$ for fixed propagation direction at $\vartheta=\pi/2$. The blue, orange, green, red and purple curves (from top to bottom) correspond to $B/T^{2}=\{0,2.01,4.22,8.73,11.24\}$ respectively.}
\label{Gamma_Out_Pi2}
\end{figure}

Once we have obtained the spectral function $\chi_{out}$, the differential rate of emitted photons can be computed from \eqref{diffproduction} integrating $\vec{k}$ over the reaction plane as we did for the in-plane polarization state \eqref{GammaInt1}. In Fig. \ref{Gamma_Out_bmax} we plot this quantity as a function of $\omega$ for a fixed value of the magnetic field $B/T^{2}=11.24$ and different values of $\vartheta$.  We can see that for any given frequency the rate of emitted photons increases as the direction of propagation aligns with the reaction plane. In Fig. \ref{Gamma_Out_Pi2}, we plot $d\Gamma_{out}/(dk_{0}d\cos\vartheta)$ as a function of the frequency for a fixed propagation direction $\vartheta=\pi/2$ and different values of $B/T^{2}$. It can be seen that the rate of emitted photons decreases as the magnetic field intensity increases for all values of $\omega$. This behavior is different from what we showed for the in-plane polarization state, where the MC or IMC phenomena was present depending on the photon frequency, while in the case of the out-plane polarization state we have only the IMC phenomenon. Once again, in spite of presenting only the results for a single magnetic field in Fig. \ref{Gamma_Out_bmax} or a single $\vartheta$ in \ref{Gamma_Out_Pi2}, we have verified that the qualitative behavior is the same for different values of $B$ or $\vartheta$.

\begin{figure}
\begin{center}
\includegraphics[width=0.4\textwidth]{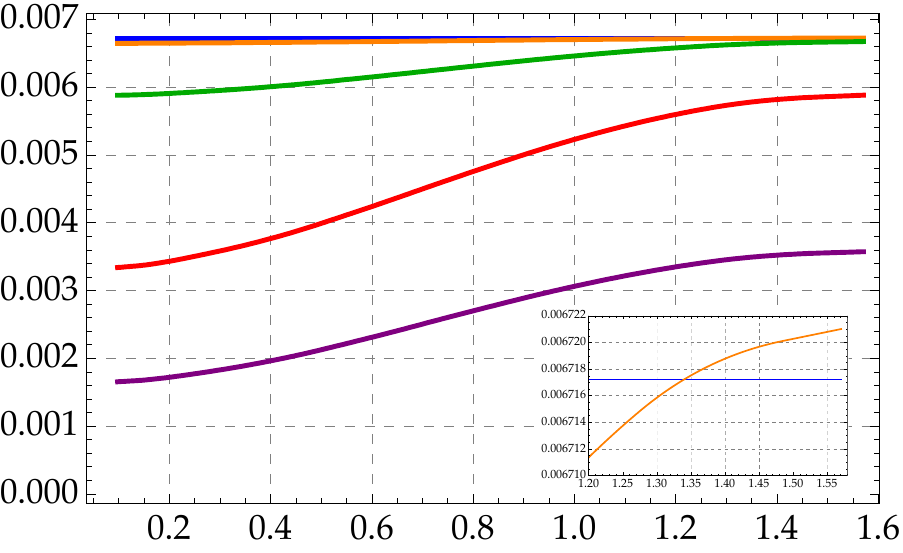} 
   \put(0,-10){$\vartheta$}
   \put(-220,130){$\frac{1}{8 N_{c}^{2}\alpha_{EM}T^{4}}\frac{d\Gamma_{out}}{d\cos\vartheta}$}
\end{center}
\caption{\small Differential photon production $\frac{d\Gamma_{out}}{d\cos\vartheta}$ for the out-plane polarization state as a function of $\vartheta$. The blue, orange, green, red and purple (from top to bottom on the left side of the graph) correspond to $B/T^{2}=\{0,2.01,4.22,8.73,11.24\}$ respectively.}
\label{Gamma_Out_Theta}
\end{figure}

Next we integrate $\omega$ in \eqref{GammaInt1} to obtain the differential photon production for the out-plane polarization state at a given direction and fixed magnetic field. In FIG. \ref{Gamma_Out_Theta} we show this quantity as a function of the propagation angle $\vartheta$ for different magnetic field intensities. We can see that for any $B$ the production of photons increases as the momentum is aligned with the reaction plane, and also that for strong magnetic field (for instance $B/T^{2}=11.24$) less photons are produced compared to the $B=0$ case. However, for a sufficiently weak magnetic field, for instance $B/T^{2}=2.01$, as it can be observed in FIG. \ref{Gamma_Out_Theta}, this is no longer the case, as for $\vartheta>1.35$ more photons are produced when compared to the $B=0$ case, as can be seen in the insert of the graphic.

\begin{figure}
\begin{center}
\includegraphics[width=0.4\textwidth]{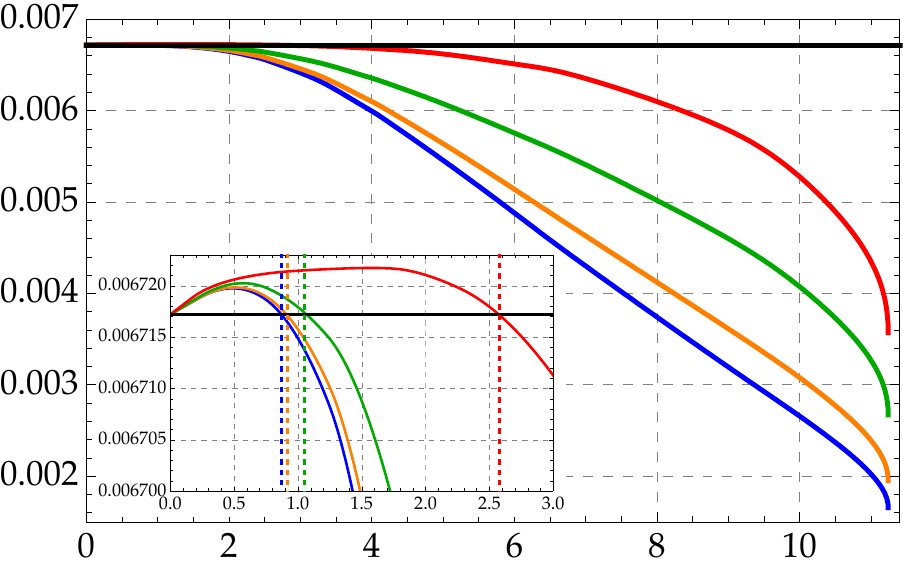} 
   \put(0,-10){$\frac{B}{T^{2}}$}
   \put(-220,140){$\frac{1}{8 N_{c}^{2}\alpha_{EM}T^{4}}\frac{d\Gamma_{out}}{d\cos\vartheta}$}
\end{center}
\caption{\small Differential photon production $\frac{d\Gamma_{out}}{dk_{0}d\cos\vartheta}$ for the out-plane polarization state as a function of $B/T^{2}$. The blue, orange, green and red curves (from bottom to top on the right side of the graph) correspond to $\vartheta=\{\pi/32, \pi/8, \pi/4, \pi/2\}$, respectively. The horizontal black line represents the total photon production for $B/T^{2}=0$. The vertical dashed lines correspond to the $B^{out}_{\vartheta}$ magnetic field intensities below which the production of photons is enhanced with respect to the $B=0$ case. For the propagation angles considered we have $B^{out}_{\pi/2}/T^{2}=2.58$, $B^{out}_{\pi/4}/T^{2}=1.05$, $B^{out}_{\pi/8}/T^{2}=0.92$ and $B^{out}_{\pi/32}/T^{2}=0.87$.}
\label{Gamma_Out_b}
\end{figure}

The behavior we  have just described, which qualitatively coincides with what we found for the in-plane state, can be better appreciated in FIG. \ref{Gamma_Out_b}, where we show the differential photon production as a function of the magnetic field for different directions of propagation. It can be seen that for any $B/T^{2}$ the production of photons is reduced as the $\vartheta\rightarrow 0$, that is, as the direction of propagation coincides with the magnetic field one. However, for any given $\vartheta$ we see the same phenomenon that we found for the in-plane polarization: there exists a magnetic field intensity $B^{out}_{\vartheta}$ such as for $B>B^{out}_{\vartheta}$ the production of photons is decreased when compared to the $B=0$ case, while for $0<B<B^{out}_{\vartheta}$ the behavior is reversed and more photons are produced when compared to the $B=0$ case. Some values for $B^{out}_{\vartheta}$ are shown in FIG. \ref{Gamma_Out_b} as vertical dotted lines. Up to our numerical precision, we were able to corroborate that $B^{out}_{\vartheta}\rightarrow 0$ as $\vartheta\rightarrow 0$, and also that $B^{out}_{\vartheta}<B^{in}_{\vartheta}$ for the explored values of $\vartheta$.

\section{Total photon production}
\label{Total}
\begin{figure}
\begin{center}
\includegraphics[width=0.4\textwidth]{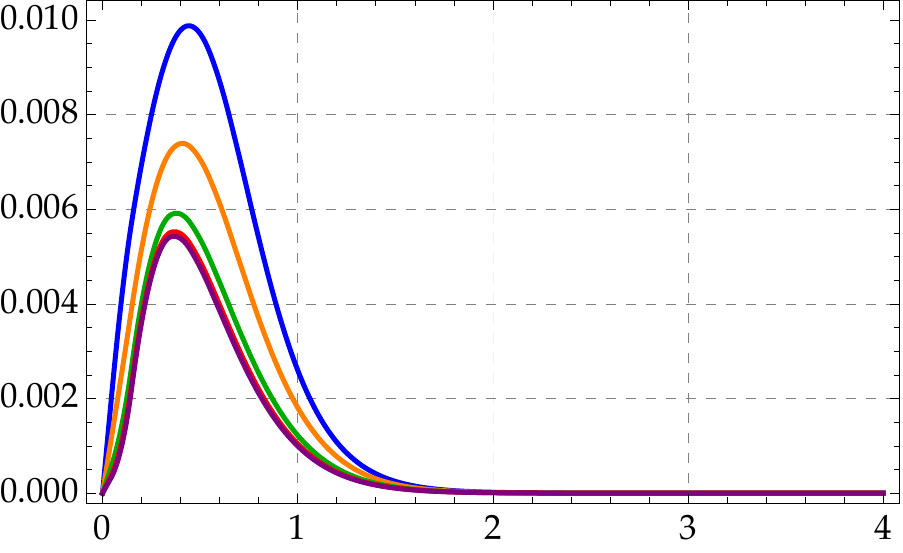} 
   \put(0,-10){$\omega$}
   \put(-220,140){$\frac{1}{8 N_{c}^{2}\alpha_{EM}T^{3}}\frac{d\Gamma}{dk_{0}d\cos\vartheta}$}
\end{center}
\caption{\small Differential photon production $\frac{d\Gamma}{dk_{0}d\cos\vartheta}$ as a function of the photon frequency $\omega=k_{0}/2\pi T$ for fixed magnetic field at $B/T^{2}=11.24$. The blue, orange, green, red and purple curves (from top to bottom) correspond to $\vartheta=\{\pi/2, \pi/4, \pi/8, \pi/16, \pi/32\}$ respectively.}
\label{Gamma_Total_bmax}
\end{figure}

\begin{figure}
\begin{center}
\includegraphics[width=0.4\textwidth]{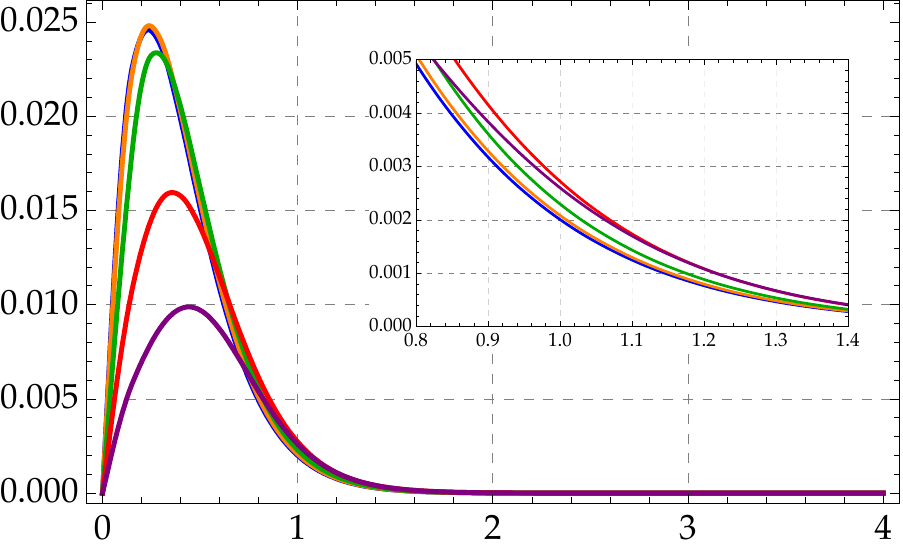} 
   \put(0,-10){$\omega$}
   \put(-220,130){$\frac{1}{8 N_{c}^{2}\alpha_{EM}T^{3}}\frac{d\Gamma}{dk_{0}d\cos\vartheta}$}
\end{center}
\caption{\small Differential photon production $\frac{d\Gamma}{dk_{0}d\cos\vartheta}$ for the out-plane polarization state as a function of the photon frequency $\omega=k_{0}/2\pi T$ for fixed propagation direction at $\vartheta=\pi/2$. The blue, orange, green, red and purple curves (from top to bottom) correspond to $B/T^{2}=\{0,2.01,4.22,8.73,11.24\}$ respectively.}
\label{Gamma_Total_Pi2}
\end{figure}

Finally, in this Section, we show the results for the total photon production \eqref{GammaTotal}, that is, considering the contributions of both polarization states. In FIG. \ref{Gamma_Total_bmax} we show $d\Gamma/(dk_{0}d\cos\vartheta)$ as a function of the photon frequency for fixed $B/T^{2}=11.24$ and different values of $\vartheta$. It can be seen that for all $\omega$ the production of photons is reduced as $\vartheta\rightarrow 0$. Comparing qualitatively this result to what we obtained previously for each polarization separately, we see that in this case the behavior is the same as the out-plane case (we explicitly checked that this is true for other values of $B/T^{2}$). On the other hand, as displayed in FIG. \ref{Gamma_Total_Pi2}, for fixed $\vartheta=\pi/2$ and small $\omega$ the production of photons decreases as the magnetic field intensity is increased (IMC), while for large $\omega$ the opposite is true (MC). This coincides with what we previously obtained for the in-plane polarization state. 

\begin{figure}
\begin{center}
\includegraphics[width=0.4\textwidth]{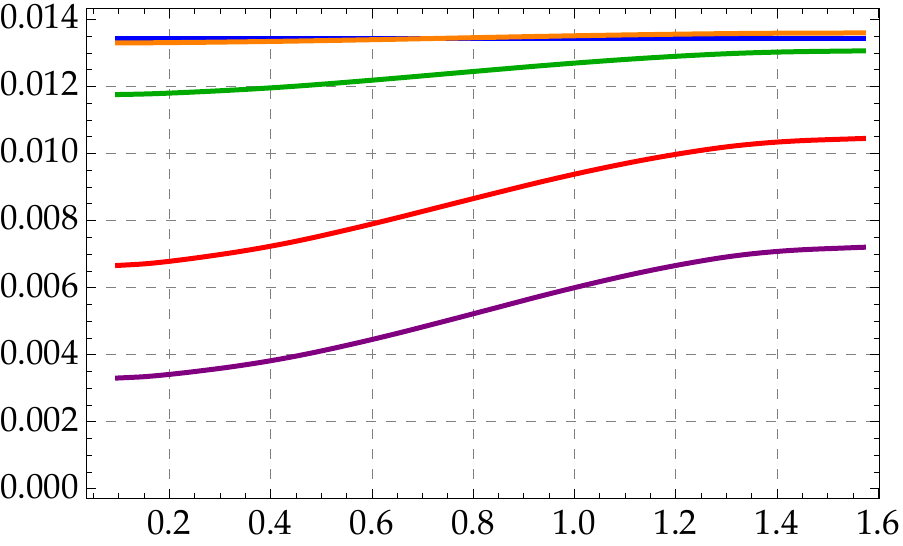} 
   \put(0,-10){$\vartheta$}
   \put(-220,130){$\frac{1}{8 N_{c}^{2}\alpha_{EM}T^{4}}\frac{d\Gamma}{d\cos\vartheta}$}
\end{center}
\caption{\small Differential photon production $\frac{d\Gamma}{d\cos\vartheta}$ as a function of $\vartheta$. The blue, orange, green, red and purple (from top to bottom on the left side of the graph) correspond to $B/T^{2}=\{0,2.01,4.22,8.73,11.24\}$ respectively.}
\label{Gamma_Total_Theta}
\end{figure}

Next, we analyze $d\Gamma/d\cos\vartheta$ after integrating the photon frequency. In FIG. \ref{Gamma_Total_Theta} we plot this quantity as a function of the propagation angle $\vartheta$ for different values of the magnetic field. The behavior is the same as the one we observed for each polarization state individually: for small $\vartheta$ increasing the magnetic field results in a decrease in the production of photons, while for sufficiently large $\vartheta$ there exist a window where increasing the magnetic field also increases the production of photons. For instance, in FIG. \ref{Gamma_Total_Theta} for fixed $B/T^{2}=2.01$ and $\vartheta<\pi/4$ less photons are produced when compared to the $B=0$ case (the black line in the plot), but for $\vartheta>\pi/4$ the opposite is true.

\begin{figure}
\begin{center}
\includegraphics[width=0.4\textwidth]{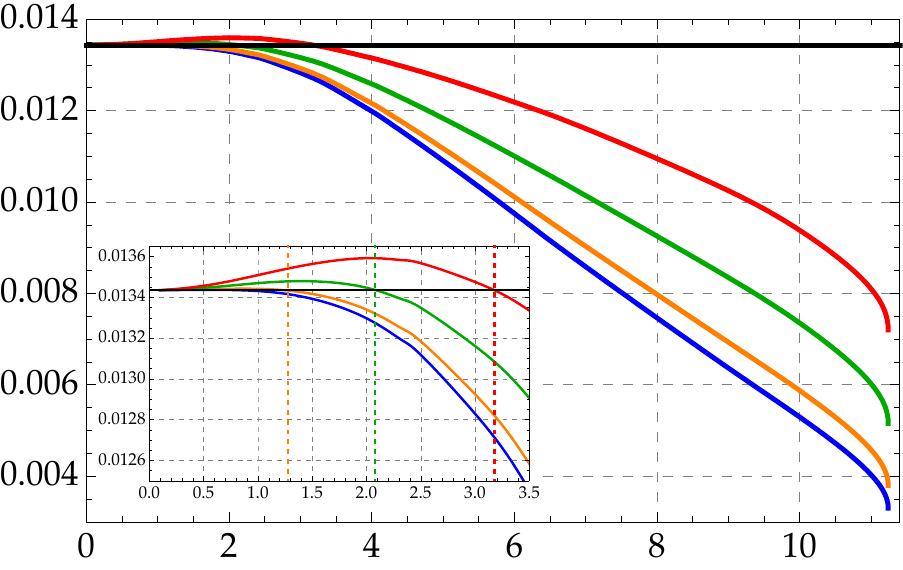} 
   \put(0,-10){$\frac{B}{T^{2}}$}
   \put(-220,140){$\frac{1}{8 N_{c}^{2}\alpha_{EM}T^{4}}\frac{d\Gamma}{d\cos\vartheta}$}
\end{center}
\caption{\small Differential photon production $\frac{d\Gamma}{dk_{0}d\cos\vartheta}$ as a function of $B/T^{2}$. The blue, orange, green and red curves (from bottom to top on the right side of the graph) correspond to $\vartheta=\{\pi/32, \pi/8, \pi/4, \pi/2\}$, respectively. The horizontal black line represents the total photon production for $B/T^{2}=0$. The vertical dashed lines correspond to the $B_{\vartheta}$ magnetic field intensities below which the production of photons is enhanced with respect to the $B=0$ case. For the propagation angles considered we have $B_{\pi/2}/T^{2}=3.18$, $B_{\pi/4}/T^{2}=2.08$ and $B_{\pi/8}/T^{2}=1.28$.}
\label{Gamma_Total_b}
\end{figure}

This phenomenon can be better appreciated in FIG. \ref{Gamma_Total_b}, where we show $d\Gamma/d\cos\vartheta$ as a function of the magnetic field for different values of $\vartheta$, representing them as curves of different colors. It can be seen that, for each value of $\vartheta$ there exists a magnetic field intensity $B_{\vartheta}$ that is a threshold for the production of photons with respect to the vanishing magnetic field case, that is, for $B<B_{\vartheta}$ more photons are produced with respect to the $B=0$ case, while for $B_{\vartheta}<B$ the opposite is observed. To our numerical accuracy we were able to find that $B_{\vartheta}\rightarrow 0$ as $\vartheta\rightarrow 0$. We were also able to corroborate that for a given $\vartheta$ we have that $B^{out}_{\vartheta}<B_{\vartheta}<B^{in}_{\vartheta}$. 

\section{Elliptic Flow}
\label{EllipticFlow}
\begin{figure}
\begin{center}
\includegraphics[width=0.4\textwidth]{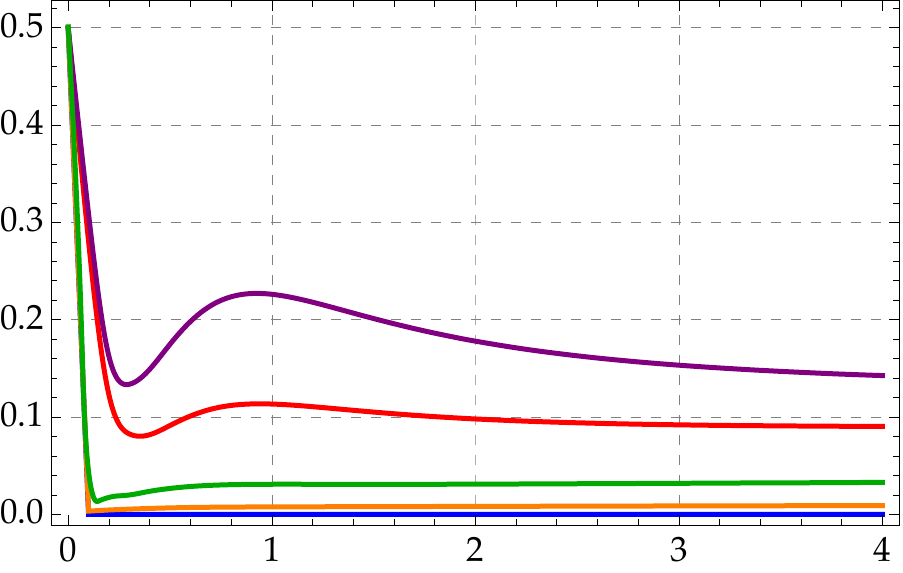} 
   \put(0,-10){$\omega$}
   \put(-205,135){$v_{2}$}
\end{center}
\caption{\small Magnetic contribution to the elliptic flow $v_{2}$ as a function of $\omega$. The blue, orange, green, red and purple curves (from bottom to top at the right) correspond to $B/T^{2}=\{0.1,2.01,4.22,8.73,11.24\}$ respectively.}
\label{v2Plot}
\end{figure}

In this section, we present in FIG. \ref{v2Plot} the magnetic contribution to the elliptic flow \eqref{v2} as a function of the photon frequency for various field intensities.

It can be seen that the general effect of the magnetic field is to increase the value of $v_{2}$ for all $\omega\neq 0$. This result suggests that the magnetic field generated in the collision could be responsible for the results reported in \cite{PHENIX:2011oxq, Lohner:2012ct} or at least play a significant role in the measured $v_2$ for thermal photons at low transverse momentum. Given that the magnetic field is the only source of anisotropy, it can be observed that as $B \rightarrow 0$ we have a lower value of component $v_{2}$ for all $\omega\neq 0$. The only exception for this behavior is in the limit $\omega\rightarrow 0$, in which the value of $v_{2}$ approaches $1/2$ regardless of the intensity of $B$ just as long as it is not vanishing.

The universal behavior just highlighted can be understood noticing from \eqref{FourierExpansion} that $v_{2}=1/2$ is the maximum value that the elliptic flow can take, as in this case $\cos(2\theta)$ is the only component in the azimuthal expansion. This saturation in the $\omega\rightarrow 0$ limit is an indication that the effect of any non-vanishing magnetic field on the ellipticity of the flow becomes overwhelming for photons with extremely low energy, response that is well captured by the dimensionless quantity $B/\omega^2$, that diverges as the frequency goes to zero, unless $B=0$.

\begin{figure}
\begin{center}
\includegraphics[width=0.4\textwidth]{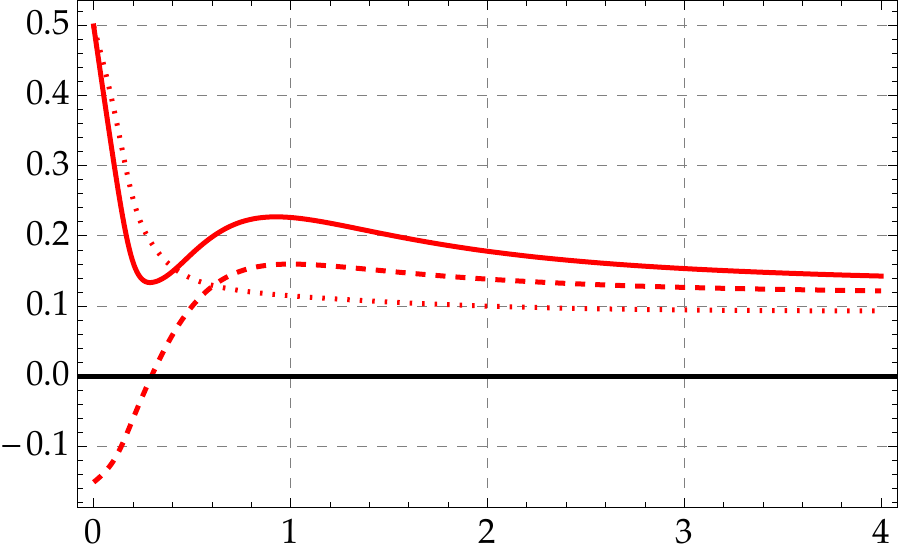} 
   \put(0,-10){$\omega$}
   \put(-205,135){$v_{2}$}
\end{center}
\caption{\small Magnetic contribution to the elliptic flow $v_{2}$ as a function of $\omega$ for fixed $B/T^{2}=8.73$. The dashed and dotted curves corresponds to the in-plane and out-plane polarization states respectively. The continuous curve correspond to the one from the averaged emission rate of two types of polarizations.}
\label{v2Plot2}
\end{figure}

In FIG. \ref{v2Plot2} we display the elliptic flow distinguishing the contribution of each of the polarization states separately and the full result for fixed $B/T^{2}=8.73$. The dashed line corresponds to the in-plane state, the dotted one to the out-plane state and the continuous line to the result from both contributions. It can be seen that for high frequencies $v_{2}$ is larger for the in-plane polarization state than for the out-plane one. However, for sufficiently small $\omega$ the elliptic flow is larger for the out-plane state.

\section{Discussion}
\label{Discussion}
In this paper we employed holographic methods to study the effect that an external magnetic field has in the production of photons in a strongly coupled plasma. While there have been other studies on this topic in the past \cite{Mamo:2013efa,Arciniega:2013dqa,Wu:2013qja,Yee:2013qma,Muller:2013ila}, our holographic model features many novel results. In particular, an important difference of our model in comparison to the one considered in \cite{Arciniega:2013dqa} is that we feature a bulk scalar field. We will discuss those differences below.

We show the numerical results for the production of photons for the in-plane (SEC. \ref{InPlane}) and the out-plane (SEC. \ref{OutPlane}) polarization states independently and the total production taking into account the contributions coming from both (SEC. \ref{Total}). In general, we find that whether a polarization state is preferred over the other depends on the intensity of the magnetic field. For instance, from FIG. \ref{Gamma_b} and \ref{Gamma_Out_b} it can be seen that for $B>B^{in}_{\vartheta}$ more photons are produced in the out-plane polarization state that in the in-plane one, in agreement with \cite{Wu:2013qja}. However, for $B<B^{in}_{\vartheta}$ the opposite effect occurs: more photons are produced in the in-plane polarization state. Other works have reported that the presence of an external magnetic field (or other spatial anisotropies) can cause an increase in the production of photons in one polarization state over the other \cite{Patino:2012py,Yee:2013qma,Arciniega:2013dqa,Wu:2013qja}.

From our results, we conclude that for any fixed non-zero magnetic field, the production is increased when the photon momentum is aligned with the reaction plane. This differs from the results from \cite{Patino:2012py}, where the emission of photons propagating along the anisotropic direction is increased, indicating that the source of the anisotropy is very relevant when studying the emission of photons, nonetheless, our results agree with previous models with a magnetic field. By introducing an external magnetic field in the model as in \cite{Patino:2012py}, the authors in \cite{Wu:2013qja} found an enhancement in the plane perpendicular to the magnetic field. Interestingly, they also found that this enhancement is more notorious for the out-plane state, meaning that using their holographic model it can be concluded that more photons are produced in its state perpendicular to the reaction plane. In our model, as described above, the effect seems to be more subtle.

As previously stated, a study of the production of photons in the presence of a magnetic field was performed in \cite{Arciniega:2013dqa} using the holographic model from \cite{DHoker:2009mmn}. In that work, due to the intricacy of the equations of motion, two directions of propagation were considered only, namely parallel and perpendicular to the magnetic field. In that sense, here we present an extension to that work, as we were able to explore the emission of photons in any propagation direction. In \cite{Arciniega:2013dqa} it was previously discovered that the spectral function for the in-plane polarization state $\chi_{in}$ dropped to zero at $\omega=0$ for any $B/T^{2}$ along parallel and perpendicular directions with respect to the magnetic field. Here we confirmed that this is indeed the case for any direction of propagation, as can be seen in FIG. \ref{Chi_Pi4}. We also show that this is not true for the out-plane polarization state $\chi_{out}$ in FIG. \ref{Chi_Out_Pi4}, where it can be seen that the curves tend to a constant finite value as $\omega\rightarrow 0$ instead.

Another main result of our study is that for any given direction of propagation, the production of photons is increased with respect to the $B=0$ value if the magnetic field intensity lies in the interval $0<B<B_{\vartheta}$, while it decreases for any $B>B_{\vartheta}$. The magnetic field intensity $B_{\vartheta}$ depends on the propagation direction in a manner such that it tends to zero as the photon momentum is aligned with the magnetic field. This effect is true regardless of the polarization state, although the magnetic field intensity at which happens does indeed depends on the polarization, that is, $B^{out}_{\vartheta}<B^{in}_{\vartheta}$ for any given $\vartheta$. This is a novelty of the holographic model studied in this work.

We also computed the magnetic contribution to the elliptic flow $v_{2}$ for the emitted photons. This flow characterizes the momentum anisotropy of the produced particles in heavy-ion collisions. The effect that an external magnetic field has $v_2$ was previously computed holographically in \cite{Muller:2013ila}. In their model the magnetic field was introduced as an excitation over probe D7-branes. Considering massless quarks, the authors found that for small frequencies $\omega$ the elliptic flow is bigger for the out-plane polarization state, while for higher frequencies the opposite behavior is displayed and the flow is bigger for the in-plane polarization states. This behavior coincides with our results in FIG. \ref{v2Plot2}. We also show in FIG. \ref{v2Plot} that the general effect of the magnetic field is to increase the value of $v_{2}$, thus it can explain the puzzling excess observed in heavy-ion experiments in RHIC and LHC \cite{PHENIX:2011oxq, Lohner:2012ct,David:2019wpt}. It was previously argued that the magnetic field was the cause of this excess by a perturbative computation in QCD in \cite{Ayala:2016lvs,Ayala:2017vex}, hence, the results presented here confirm the effect of the magnetic field on $v_2$ using a non-perturbative calculation.  

There are two further comments that might be relevant concerning the applicability of what we have concluded to a real world quark-gluon plasma. One is that the magnetic field created in a collision experiment is expected to have a lifetime of a fraction of a $fm/c$ even at top energy in either RHIC or LHC. During this time the temperature is also very high, so a careful determination of the ratio $B/T^2$ is necessary to do any comparison with our plots.  A second comment is that a word of caution has been put forward in \cite{Wang:2021oqq}, where the authors indicate that if the time it takes for Ohm's law to be applicable is longer than the above mentioned lifetime, the electromagnetic response would be incomplete, and the maximum intensity of the magnetic field could be diminished by up to two orders of magnitude. If this is indeed the case, the experiments could be in a range of $B/T^2$ where the effects we have predicted are difficult to observe.

We would like to close commenting that the results presented in this manuscript constitute a completion of the ones we previously showed in \cite{Avila:2021rcu}, as all the results for the in-plane polarization state apply to both setups. The reason is that the fully back-reacted equations involving the in-plane polarization photons are identical to those we found in \cite{Avila:2021rcu}. A generalization of our work, which we will present elsewhere \cite{Avila2come}, is to return to the setting of \cite{Avila:2021rcu} and consider a full treatment of the perturbations in the truncated theory.

\section*{Acknowledgments}
LP and FN acknowledge partial financial support from PAPIIT-UNAM IN113618. DA is partially supported by Mexico's National Council of Science and Technology (CONACyT) grant A1-S-22886 and DGAPA-UNAM grants IN107520 and IN116823 and additionally supported by by CONACYT Ph.D. grant CVU 745912 and DGAPA-UNAM postdoctoral grant. 
\\

\appendix
\section{Equations of motion for the in-plane polarization state}
\label{AppIn}
For the polarization state $\epsilon_{in}$ we need to solve thirteen differential equations for nine components of the fields: $A_{y}$, $\phi$, $h_{tx}$, $h_{tz}$, $h_{xz}$, $h_{tt}$, $h_{xx}$, $h_{yy}$ y $h_{zz}$. Even when the number of equations surpasses the number of variables, the system can be solved consistently by the procedure described in the main text. The metric fields $h_{mn}$ are the rescaled versions given in \eqref{rescaling}, while the $\log(r)$ has yet to be factorized from the scalar field $\phi$. The reason for this is that numerically is more efficient to solve for this variable $\phi$ and then divide by $\log(r)$ at the end of the computation. 

The thirteen equations are
\begin{widetext}
\begin{eqnarray}
&&
\begin{aligned}
0=&V(8i\sqrt{6}B k_{0}r^{4}A_{y}\sin\vartheta UW^{2}-rUW(8B^{2}rW\phi-e^{\frac{\varphi}{\sqrt{6}}}V^{2}(2\phi(W(-6e^{\frac{\varphi}{\sqrt{6}}}U'+8re^{\sqrt{\frac{3}{2}}\varphi}+4r)-3k_{0}^{2}r\cos^{2}\vartheta e^{\frac{\varphi}{\sqrt{6}}})\\&-3re^{\frac{\varphi}{\sqrt{6}}}W(r^{3}(rh'_{tt}+2h_{tt})\varphi'-2U'\phi'))+6k_{0}^{2}r\sin^{2}\vartheta e^{\sqrt{\frac{2}{3}}\varphi}VW\phi)+3r^{2}e^{\sqrt{\frac{2}{3}}\varphi}V^{2}W^{2}(r^{4}h_{tt}U'\varphi'+2k_{0}^{2}\phi)\\&+3U^{2}e^{\sqrt{\frac{2}{3}}\varphi}V(rW^{2}(r^{5}\varphi'(h'_{xx}+h_{yy}')+2rV'\phi'- 4\phi V')+V(rW(r^{4}(rh'_{zz}+2h_{zz})\varphi'+ rW'\phi'- 2\phi W')\\&-r^{6}h_{zz}\varphi' W'+2W^{2}(r(r\phi''-4\phi')+6\phi))))-r^{5}h_{xx}UW^{2} (4\sqrt{6}B^{2}r-3Ue^{\sqrt{\frac{2}{3}}\varphi}V\varphi'(2V-rV'))\\&-r^{5}h_{yy}UW^{2}(4\sqrt{6}B^{2}r-3Ue^{\sqrt{\frac{2}{3}}\varphi}V\varphi'(2V-rV'))
\end{aligned}
\cr
&&
\cr
&&
\begin{aligned}
0=&3k_{0}^{2}r^{2}VA_{y}(U(V-W)\cos(2 \vartheta )+U (V+W)-2 V W)+r^2 U V^2 (U (-3 W' A_y'+2 \sqrt{6} W \varphi ' A_y'-6 W A_y'')\\&-6 W U' A_y')+i Bk_0 (3 r^4 V W h_{tt} \sin\vartheta +6 r^4 V W h_{tx}+U (\sin\vartheta(-3 r^4 V h_{zz}+3 r^4 W h_{yy}+2 \sqrt{6} V W \phi )\\&+6 r^4 V h_{xz} \cos\vartheta+3 r^4 W h_{xx} \sin\vartheta))
\end{aligned}
\cr
&&
\cr
&&
\begin{aligned}
0=&-(8e^{-\sqrt{\frac{2}{3}}\varphi}(-6iBk_{0}r^{2}VA_{y}\sin\vartheta +3B^{2}r^{4}h_{xx}+3B^{2}r^{4}h_{yy}+\sqrt{6}B^{2}V\phi -2\sqrt{6}e^{2\sqrt{\frac{2}{3}}\varphi}V^{3}\phi +2\sqrt{6}e^{\frac{\varphi}{\sqrt{6}}}V^{3}\phi))\\&(r^{2}UV^{3})^{-1}-(36\varphi'(r\phi'-2\phi)r^{-3}+U^{-3}(9(h_{tt}(-2r^{2}UU''+r^{2}U'^{2}-2rUU'+4U^{2})-(U(rVW(V^{2}W^{2}h'_{tt}(rU'\\&-8U)+U(W^{2}h'_{xx}(rVU'-2rUV'+8UV)+W^{2}h'_{yy}(rVU'-2rUV'+8UV)+V(2rW(-VWh''_{tt}+UVh''_{zz}\\&+UWh''_{xx}+UWh''_{yy})+Vh'_{zz}(rWU'-2rUW'+8UW))))+UV^{3}h_{zz}(2U(W(2W-r^{2}W'')+r^{2}W'^{2}-2rWW')\\&+rWU'(2W-rW'))+UW^{3}h_{xx}(2U(V(2V-r^{2}V'')+r^{2}V'^{2}-2rVV')+rVU'(2V-rV'))\\&+UW^{3}h_{yy}(2U(V(2V-r^{2}V'')+r^{2}V'^{2}-2rVV')+rVU'(2V-rV'))))(V^-3 W^-3)))
\end{aligned}
\cr
&&
\cr
&&
\begin{aligned}
0=&r^{3}(W(-2V(rU(W(\sin\vartheta h'_{tx}+h'_{xx}+h'_{yy})+V(\cos\vartheta h'_{tz}+h'_{zz}))+Wh_{tx}\sin\vartheta(2U-rU'))\\&+Wh_{xx}(rVU'+rUV'-4UV)+Wh_{yy}(rVU'+rUV'-4UV))-2V^{2}Wh_{tz}\cos\vartheta(2U-rU')\\&+V^{2}h_{zz}(rWU'+rUW'-4UW))(U V^2 W^2)^{-1}-2\phi\varphi'
\end{aligned}
\cr
&&
\cr
&&
\begin{aligned}
0=&-8Be^{-\sqrt{\frac{2}{3}}\varphi}A'_{y}V^{-1}-2ir^{-2}k_{0}\phi\varphi'\sin\vartheta+(ik_{0}r(-W(V(-2rU(VW(\sin\vartheta h'_{tt}+h'_{tx})+UV(\cos\vartheta h'_{xz}-\sin\vartheta h'_{zz})\\&-UW\sin\vartheta h'_{yy})+Wh_{tt}\sin\vartheta(rVU'+rUV'-4UV)+2UWh_{tx}(rV'-2V))+2U^{2}Wh_{yy}\sin\vartheta(2V-rV'))\\&+U^{2}Vh_{zz}\sin\vartheta(rWV'+rVW'-4VW)+2U^{2}VWh_{xz}\cos\vartheta(2V-rV')))(U^2 V^2 W^2)^{-1}
\end{aligned}
\cr
&&
\cr
&&
\begin{aligned}
0=&2\phi\varphi'\cos\vartheta+ r^{3}W^{-1}(U^{-2}(h_{tt}\cos\vartheta(rWU'+rUW'-4UW)-2U(rW(\cos\vartheta h'_{tt}+h'_{tz})+h_{tz}(2W-rW')))\\&+V^{-2}(2V(rW(\cos\vartheta h'_{xx}-\sin\vartheta h'_{xz})+h_{xz}\sin\vartheta(rW'-2W))-h_{xx}\cos\vartheta(rWV'+rVW'-4VW)\\&+(\cos\vartheta(2rVWh'_{yy}-h_{yy}(rWV'+rVW'-4VW))))
\end{aligned}
\cr
&&
\cr
&&
\begin{aligned}
0=&r^{-2}(8Ue^{-\sqrt{\frac{2}{3}}\varphi}(-6iBk_{0}r^{2}VA_{y}\sin\vartheta+3B^{2}r^{4}h_{xx}+3B^{2}r^{4}h_{yy}+\sqrt{6}B^{2}V\phi-2\sqrt{6}e^{2\sqrt{\frac{2}{3}}\varphi}V^{3}\phi +2\sqrt{6}e^{\frac{\varphi}{\sqrt{6}}}V^{3}\phi))\\&+24r^{2}e^{-\sqrt{\frac{2}{3}}\varphi}Vh_{tt}(B^{2}+2e^{\frac{\varphi}{\sqrt{6}}}(e^{\sqrt{\frac{3}{2}}\varphi}+2)V^{2})+(UW^{2})^{-1}(9V(rU(W(V(h'_{tt}(rVWU'-U(2rWV'+rVW'\\&+8VW))+rU(U'(Vh'_{zz}+Wh'_{xx}+Wh'_{yy})-2VWh''_{tt})+4k_{0}^{2}rWh_{tx}\sin\vartheta)+Wh_{xx}(2k_{0}^{2}rV+UU'(2V-rV'))\\&+Wh_{yy}(2k_{0}^{2}rV+UU'(2V-rV')))+4k_{0}^{2}rV^{2}Wh_{tz}\cos\vartheta +V^{2}h_{zz}(2k_{0}^{2}rW+UU'(2W-rW')))\\&+VWh_{tt}(2rU(k_{0}^{2}r(V\cos^2 \vartheta +W\sin^2 \vartheta)+VWU')-r^{2}VWU'^{2}-2U^{2}(2rWV'+rVW'+2VW))))
\end{aligned}
\cr
&&
\cr
&&
\begin{aligned}
0=&24iBk_{0}e^{-\sqrt{\frac{2}{3}}\varphi}VA_{y}+8r^{2}e^{-\sqrt{\frac{2}{3}}\varphi}h_{tx}(B^{2}+2e^{\frac{\varphi}{\sqrt{6}}}(e^{\sqrt{\frac{3}{2}}\varphi}+2)V^{2})-W^{-1}(3V(r(2rUVWh''_{tx}+rUVW'h'_{tx}\\&+8UVWh'_{tx}+k_{0}^{2}rVh_{tz}\sin(2\vartheta)-2k_{0}^{2}rVh_{xz}\cos\vartheta+ 2k_{0}^{2}rVh_{zz}\sin\vartheta+2k_{0}^{2}rWh_{yy}\sin\vartheta)-2h_{tx}(k_{0}^{2}r^{2}V\cos^2 \vartheta \\&+r^{2}WU'V'+rUVW'+2UVW)))
\end{aligned}
\cr
&&
\cr
&&
\begin{aligned}
0=&h_{tz}(8B^{2}r^{2}e^{-\sqrt{\frac{2}{3}}\varphi}+6rV(k_{0}^{2}r\sin^{2}\vartheta- 2UV')+2V^{2}(8r^{2}e^{\sqrt{\frac{2}{3}}\varphi}+16r^{2}e^{-\frac{\varphi}{\sqrt{6}}}-3rW^{-1}W'(r U'-U)-6U))\\&-W^{-1}(3rV(2rUVWh''_{tz}+2rUWV'h'_{tz}-rUVW'h'_{tz}+8UVWh'_{tz}+k_{0}^{2}rWh_{tx}\sin(2\vartheta)+2k_{0}^{2}rWh_{xx}\cos\vartheta\\&- 2k_{0}^{2}rWh_{xz}\sin\vartheta+ 2k_{0}^{2}rWh_{yy}\cos\vartheta))
\end{aligned}
\cr
&&
\cr
&&
\begin{aligned}
0=&-96iBk_{0}e^{-\sqrt{\frac{2}{3}}\varphi}VA_{y}\sin\vartheta +3rh_{yy}(16B^{2}re^{-\sqrt{\frac{2}{3}}\varphi}+6k_{0}^{2}rV\sin^{2}\vartheta+ 3rUV'^{2}-6UVV')\\&+3h_{xx}(3k_{0}^{2}r^{2}(UW)^{-1}V^{2}(U\cos(2\vartheta)+U-2W)-3r^{2}UV'^{2}+16r^{2}e^{\sqrt{\frac{2}{3}}\varphi}V^{2}+32r^{2}e^{-\frac{\varphi}{\sqrt{6}}}V^{2}-12rV^{2}U'\\&-6rUW^{-1}V^{2}W'+6rUVV'-12UV^{2})-r^{-2}(V(-16\sqrt{6}B^{2}e^{-\sqrt{\frac{2}{3}}\varphi}\phi +18r^{4}UVh''_{xx}-9r^{4}VV'h'_{tt}+18r^{4}V U'h'_{xx}\\&+9r^{4}UW^{-1}VV'h'_{zz}-9r^{4}UV'h'_{xx}+9r^{4}UV'h'_{yy}+9r^{4}UW^{-1}VW'h'_{xx}+72r^{3}UVh'_{xx}\\&+36k_{0}^{2}r^{4}U^{-1}Vh_{tx}\sin\vartheta +36k_{0}^{2}r^{4}VW^{-1}h_{xz}\sin\vartheta\cos\vartheta -18k_{0}^{2}r^{4}VW^{-1}h_{zz}\sin^{2}\vartheta \\&+9r^{3}VU^{-1}h_{tt}(2k_{0}^{2}r\sin^{2}\vartheta +V'(rU'-2U))-9r^{4}UW^{-2}Vh_{zz}V'W'+18r^{3}UW^{-1}Vh_{zz}V'-16\sqrt{6}e^{\sqrt{\frac{2}{3}}\varphi}V^{2}\phi \\&+16\sqrt{6}e^{-\frac{\varphi}{\sqrt{6}}}V^{2}\phi))
\end{aligned}
\cr
&&
\cr
&&
\begin{aligned}
0=&2h_{xz}(2rUW(2B^{2}r+e^{\frac{\varphi}{\sqrt{6}}}V^{2}(4r(e^{\sqrt{\frac{3}{2}}\varphi}+2)-3e^{\frac{\varphi}{\sqrt{6}}}U'))-3k_{0}^{2}r^{2}e^{\sqrt{\frac{2}{3}}\varphi}V^{2}W-3U^{2}e^{\sqrt{\frac{2}{3}}\varphi}V(rW'(rV'-V)+2VW))\\&+3V(re^{\sqrt{\frac{2}{3}}\varphi}(k_{0}^{2}rUWh_{yy}\sin(2\vartheta)-V(2rU^{2}Wh''_{xz}-rU^{2}W'h'_{xz}+2rUWU'h'_{xz}+8U^{2}Wh'_{xz}+k_{0}^{2}rWh_{tt}\sin(2\vartheta)\\&+2k_{0}^{2}rWh_{tx}\cos\vartheta +2k_{0}^{2}rWh_{tz}\sin\vartheta))-8iBk_{0}UWA_{y}\cos\vartheta)
\end{aligned}
\cr
&&
\cr
&&
\begin{aligned}
0=&V(-96iBk_{0}r^{2}UW^{2}A_{y}\sin\vartheta +UW^{2}(16\sqrt{6}\phi(B^{2}+e^{\frac{\varphi}{\sqrt{6}}}(e^{\sqrt{\frac{3}{2}}\varphi}-1)V^{2})-9r^{3}e^{\sqrt{\frac{2}{3}}\varphi}V(2rU'h'_{yy}-V'(rh'_{tt}+2h_{tt})))\\&-9r^{3}U^{2}e^{\sqrt{\frac{2}{3}}\varphi}(2rVW^{2}h''_{yy}+V'(rW(Vh'_{zz}+Wh'_{xx})+Vh_{zz}(2W-rW'))+Wh'_{yy}(-rWV'+rVW'+8VW))\\&-9r^{4}e^{\sqrt{\frac{2}{3}}\varphi}VW^{2}h_{tt}U'V')+3r^{3}UW^{2}h_{xx}(16B^{2}r+3Ue^{\sqrt{\frac{2}{3}}\varphi}V'(rV'-2V))+3r^{2}e^{\frac{\varphi}{\sqrt{6}}}Wh_{yy}(-6k_{0}^{2}r^{2}e^{\frac{\varphi}{\sqrt{6}}}V^{2}W\\&+2rUV(3k_{0}^{2}re^{\frac{\varphi}{\sqrt{6}}}(V\cos^{2}\vartheta +W\sin^{2}\vartheta)+8r(e^{\sqrt{\frac{3}{2}}\varphi}+2)VW-6e^{\frac{\varphi}{\sqrt{6}}}VWU')\\&-3U^{2}e^{\frac{\varphi}{\sqrt{6}}}(r^{2}WV'^{2}+2V^{2}(rW'+2W)-2rVWV'))
\end{aligned}
\cr
&&
\cr
&&
\begin{aligned}
0=&-r^{-2}(8e^{-\sqrt{\frac{2}{3}}\varphi}W(-6iBk_{0}r^{2}VA_{y}\sin\vartheta +3B^{2}r^{4}h_{xx}+3B^{2}r^{4}h_{yy}+\sqrt{6}B^{2}V\phi -2\sqrt{6}e^{2\sqrt{\frac{2}{3}}\varphi}V^{3}\phi +2\sqrt{6}e^{\frac{\varphi}{\sqrt{6}}}V^{3}\phi))\\&+24r^{2}e^{-\sqrt{\frac{2}{3}}\varphi}Vh_{zz}(B^{2}+2e^{\frac{\varphi}{\sqrt{6}}}(e^{\sqrt{\frac{3}{2}}\varphi}+2)V^{2})+9V(-(U W^2)^{-1}(V(rW(U(2h'_{zz}(rVWU'+U(rWV'-rVW'\\&+4VW))+rW(2UVh''_{zz}+W(U(h'_{xx}+h'_{yy})-Vh'_{tt}))+2k_{0}^{2}rWh_{xz}\sin(2\vartheta))+VWh_{tt}(2k_{0}^{2}r\cos^{2}\vartheta \\&+W'(rU'-2U))+4k_{0}^{2}rVWh_{tz}\cos\vartheta)+h_{zz}(k_{0}^{2}r^{2}W^{2}(U\cos(2\vartheta)-U+2V)\\&+U(4rVW^{2}U'+U(4rW^{2}V'+V(rW'-2W)^{2})))))+rh_{xx}(2k_{0}^{2}rV\cos^{2}\vartheta +UW'(rV'-2V))\\&+rh_{yy}(2k_{0}^{2}rV\cos^{2}\vartheta +UW'(rV'-2V)))
\end{aligned}
\end{eqnarray}
\end{widetext}
\section{Equations of motion for the out-plane polarization state}
\label{AppOut}
As explained in the main text, in order to compute the production of photons in the polarization state $\epsilon_{out}$, we need to solve eight equations for the field's components $A_{t}$, $A_{x}$, $A_{z}$, $h_{ty}$, $h_{xy}$ and $h_{yz}$. Note that the metric functions $h_{mn}$ here are the rescaled versions described in \eqref{rescaling}. The eight equations are
\begin{widetext}
\begin{eqnarray}
&&
\begin{aligned}
0=U W A'_{x}\sin\vartheta+V(W A'_{t}+U A'_{z}\cos\vartheta),
\end{aligned}
\cr
&&
\cr
&&
\begin{aligned}
0=&-6UV^{2}W A''_{t}-6 UVW V'A_{t}'-3UV^{2}A_{t}'W'+2\sqrt{6}UV^{2}WA_{t}'\varphi'+6k_{0}^{2}A_{t}V (V\cos^{2}\vartheta+W\sin^{2}\vartheta)\\&+6k_{0}^{2}A_{x}VW\sin\vartheta+6k_{0}^{2}A_{z}V^{2}\cos\vartheta+6iB k_{0}r^{2}h_{ty}W\sin\vartheta,
\end{aligned}
\cr
&&
\cr
&&
\begin{aligned}
0=&6k_{0}^{2}VWA_{t}\sin\vartheta +6U^{2}VW A''_{x}+6UVW A_{x}'U'+3U^{2}V A_{x}'W'-2\sqrt{6}U^{2}VW A_{x}'\varphi' -6k_{0}^{2}A_{x}V(U\cos^{2}\vartheta -W)\\&+6 k_{0}^{2} A_{z}UV\sin\vartheta \cos\vartheta +6iBk_{0}r^{2}h_{ty}W+6iBk_{0}r^{2}\cos\vartheta h_{yz}U,
\end{aligned}
\cr
&&
\cr
&&
\begin{aligned}
0=&V(6k_{0}^{2}A_{t}\cos\vartheta VW+6k_{0}^{2}A_{x}\sin\vartheta \cos\vartheta UW+6U^{2}VWA_{z}''+6UVWA_{z}'U'+6U^{2}WA_{z}'V'-3U^{2}VA_{z}'W'\\&-2\sqrt{6}U^{2}VWA_{z}'\varphi' -6k_{0}^{2}A_{z}W(\sin^{2}\vartheta U-V))-6iBk_{0}r^{2}\sin\theta h_{yz}UW,
\end{aligned}
\cr
&&
\cr
&&
\begin{aligned}
0=& 4Be^{-\sqrt{\frac{2}{3}}\varphi}VA_{x}'+\frac{ik_{0}r}{UW}(V(rV(W h_{ty}'+\cos\vartheta U h_{yz}')+h_{ty}W(2V-rV')+r\sin\vartheta UWh_{xy}')+\sin\vartheta h_{xy}UW(2V-rV')\\&+\cos\vartheta h_{yz}UV(2V-rV')),
\end{aligned}
\cr
&&
\cr
&&
\begin{aligned}
0=&-24iBk_{0}e^{-\sqrt{\frac{2}{3}}\varphi}V(A_{t}\sin\vartheta +A_{x})+8r^{2}h_{ty}e^{-\sqrt{\frac{2}{3}}\varphi}(B^{2}+2e^{\frac{\varphi}{\sqrt{6}}}(e^{\sqrt{\frac{3}{2}}\varphi}+2)V^{2})-\frac{3V}{W}(r(UV(2rWh_{ty}''+r h_{ty}'W'\\&+8Wh_{ty}')-2k_{0}^{2}r\sin\vartheta h_{xy}W-2k_{0}^{2}r\cos\vartheta h_{yz}V)+h_{ty}(2r^{2}W(U'V'-k_{0}^{2}\sin^{2}\vartheta)+V(2U(rW'+2W)\\&-2k_{0}^{2}r^{2}\cos^{2}\vartheta)))
\end{aligned}
\cr
&&
\cr
&&
\begin{aligned}
0=& 2h_{xy}(rU(e^{\frac{\varphi}{\sqrt{6}}}V^{2}(3k_{0}^{2}r\cos^{2}\vartheta e^{\frac{\varphi}{\sqrt{6}}}-6e^{\frac{\varphi}{\sqrt{6}}}WU'+8r(e^{\sqrt{\frac{3}{2}}\varphi}+2)W(r))-8B^{2}rW)-3k_{0}^{2}r^{2}e^{\sqrt{\frac{2}{3}}\varphi}V^{2}W\\&-3U^{2}e^{\sqrt{\frac{2}{3}}\varphi}(r^{2}WV'^{2}-2rVWV'+V^{2}(rW'+2W)))-3re^{\sqrt{\frac{2}{3}}\varphi}V(2k_{0}^{2}r\sin\vartheta h_{ty}VW\\&+U(2rVW h_{xy}'U'+U(V(2rW h_{xy}''+rh_{xy}'W'+8W h_{xy}')-2rWh_{xy}'V'))+k_{0}^{2}r\sin(2\vartheta)h_{yz}UV),
\end{aligned}
\cr
&&
\cr
&&
\begin{aligned}
0=& 24iBk_{0}e^{-\sqrt{\frac{2}{3}}\varphi}V(A_{x}\cos\vartheta -A_{z}\sin\vartheta)+8r^{2}h_{yz}e^{-\sqrt{\frac{2}{3}}\varphi}(B^{2}+2e^{\frac{\varphi}{\sqrt{6}}}(e^{\sqrt{\frac{3}{2}}\varphi}+2) V^{2})-\frac{3V}{UW}(r(V(2k_{0}^{2}r\cos\vartheta h_{ty}W\\&+U(2rW h_{yz}'U'+U(2rW h_{yz}''-r h_{yz}'W'+8W h_{yz}')))+k_{0}^{2}r\sin(2\vartheta)h_{xy}UW)+h_{yz}(2k_{0}^{2}r^{2}VW\\&-2rUW(k_{0}^{2}r\sin^{2}\vartheta -2VU')+2U^{2}(r^{2}V'W'-r VW'+2VW))),
\end{aligned}
\end{eqnarray}
\end{widetext}

\newpage
\bibliography{qgp-references.bib}
\bibliographystyle{unsrt}
\end{document}